\documentclass[a4paper,fleqn,usenatbib]{mnras}

\usepackage{newtxtext,newtxmath}

\usepackage[T1]{fontenc}
\usepackage{ae,aecompl}

\usepackage{pdflscape}	

\usepackage{graphicx}	
\usepackage{amsmath}	
\usepackage{amssymb}	
\usepackage{multirow}

\usepackage{etoolbox}
\makeatletter
\patchcmd\@combinedblfloats{\box\@outputbox}{\unvbox\@outputbox}{}{\errmessage{\noexpand patch failed}}
\makeatother

\title[Disc RR~Lyrae stars]{Evidence for Galactic disc RR~Lyrae stars in the Solar neighbourhood}

\author[Prudil et al.]{
Z. Prudil$^{1}$\thanks{E-mail: prudilz@ari.uni-heidelberg.de}, I. D\'ek\'any$^{1}$, E. K. Grebel$^{1}$, A. Kunder$^{2}$\\
$^{1}$ Astronomisches Rechen-Institut, Zentrum f{\"u}r Astronomie der Universit{\"a}t Heidelberg, M{\"o}nchhofstr. 12-14, D-69120 Heidelberg, Germany\\
$^{2}$ Saint Martin's University, 5000 Abbey Way SE, Lacey, WA, 98503
\\
}

\date{Accepted XXX. Received YYY; in original form ZZZ}

\pubyear{2017}

\begin{document}

\label{firstpage}
\pagerange{\pageref{firstpage}--\pageref{lastpage}}
\maketitle

\begin{abstract}
We present a kinematical study of 314 RR~Lyrae stars in the solar neighbourhood using the publicly available photometric, spectroscopic, and {\it Gaia} DR2 astrometric data to explore their distribution in the Milky Way. We report an overdensity of 22 RR~Lyrae stars in the solar neighbourhood at a pericenter distance of between 5--9\,kpc from the Galactic center. Their orbital parameters and their chemistry indicate that these 22 variables share the kinematics and the [Fe/H] values of the Galactic disc, with an average metallicity and tangential velocity of [Fe/H]=$-0.60$\,dex and $v_{\theta} = 241$\,km\,s$^{-1}$, respectively. From the distribution of the Galactocentric spherical velocity components, we find that these 22 disc-like RR~Lyrae variables are not consistent with the {\it Gaia} Sausage ({\it Gaia}-Enceladus), unlike almost half of the local RR~Lyrae stars. Chemical information from the literature shows that the majority of the selected pericenter peak RR~Lyrae variables are $\alpha$-poor, a property shared by typically much younger stars in the thin disc. Using the available photometry we rule out a possible misclassification with the known classical and anomalous Cepheids. The similar kinematic, chemical, and pulsation properties of these disc RR~Lyrae stars suggest they share a common origin. In contrast, we find the RR~Lyrae stars associated with the {\it Gaia}-Enceladus based on their kinematics and chemical composition show a considerable metallicity spread in the old population ($\sim$~1\,dex). 
\end{abstract}




\section{Introduction}

Understanding the formation and evolution of spiral galaxies, the most common type of massive galaxies, is one of the main tasks of Galactic astrophysics. The Milky Way (MW) is a very useful laboratory for studies of the chemical and dynamical history of spiral galaxies, thanks to our ability to resolve individual stars. With the rise of large-scale photometric, spectroscopic, and astrometric surveys, we are now able to probe individual MW components (Galactic halo, disc, and bulge) to great depths and in enhanced detail.

\cite{Gilmore1983} found that the stellar number density distribution cannot be modeled by a single density profile as we move above or below the Galactic plane, but two density components are needed -- called the {\it thin} and the {\it thick} disc. The bimodality of the disc is also visible in the chemical distribution of [$\alpha$/Fe] vs. [Fe/H], where we can identify different stellar populations based on their chemical composition \citep[e.g.,][]{Fuhrmann1998,Bensby2003,Haywood2008}. This bimodality suggests a separation of the disc into a thin ($\alpha$-poor) and a thick ($\alpha$-rich) component. Moreover, the thin disc can be divided into an {\it inner} and {\it outer} thin disc, both with different metallicities, \citep[e.g.,][]{Haywood2013,Hayden2015,Bland-Hawthorn2019}. In addition, the inner disc (between the Sun and the Galactic center) is composed of the thick disc and metal-rich thin disc, while the outer disc consists of only the metal-poor thin disc \citep[e.g.,][]{Bensby2013,Anders2014,Haywood2015}. 

One of the modern formation scenarios suggests that the thick disc formed first from a turbulent, well-mixed gas disc and that its youngest part supposedly defined the conditions for subsequent build-up of the inner thin disc \citep[e.g.,][]{Haywood2013,Hayden2015,Bland-Hawthorn2019}. On the other hand, the outer thin disc might have formed independently of the thick disc with a time scale similar to the thin disc. The Galactic disc formation timelines range from 8 -- 13\,Gyr for the thick disc and up to 8\,Gyr\footnote{With a few stars reaching up to 10\,Gyr, see fig.~16 in \cite{Haywood2013}.} for the thin disc, which implies two prominent periods of star formation with the older one being associated with the thick disc \citep[e.g.,][]{Oswalt1996,Haywood2013,Kilic2017,Gallart2019}.

Many kinematical and chemical studies have been conducted over the years to study the structure of the Galactic thin and thick disc, especially using F and G dwarfs \citep[e.g.,][]{Bensby2003,Nordstrom2004,Bensby2014} in addition to public surveys \citep{Lee2011,Hayden2015,Guiglion2015,Bland-Hawthorn2019}. Furthermore, some of these studies tried to separate both MW components based on their kinematical properties, although the thin and thick disc have overlapping kinematical distributions, which hampers such efforts. On the other hand, the two MW components have different velocity dispersions, the thick disc exhibiting a higher velocity dispersion in all three velocity components, e.g., \cite{Casagrande2011} and \cite{Haywood2013}. Moreover, the thick disc population seems to lag behind the thin disc stars \citep[e.g.,][]{Lee2011}. 

In this study we focus on RR~Lyrae stars, which are old \citep[age $> 10$\,Gyr, e.g.,][]{Glatt2008,VandenBerg2013} helium-burning, pulsating horizontal branch variables. They can be divided into three types based on the pulsation mode: RRab -- fundamental mode, RRc -- first overtone, and RRd - double mode (fundamental and first overtone) pulsators. The RR Lyrae pulsators are often used as tracers of the properties of the old population of the Galactic halo and bulge \citep[e.g.,][]{Dekany2013,Belokurov2018RRlyr,Prudil2019OO}, or nearby galaxies \citep[e.g.,][]{Haschke2012RRLyrLMC,Haschke2012RRLyrSMC,JD2017RRlyrae}. In general, these variables are not used as probes of the Galactic disc, although they have been used successfully to put constraints on the structure of the thick disc and the mechanisms that contributed to its formation \citep{Kinemuchi2006,Kinman2009,Mateu2012,Mateu2018,Dekany2018}. Kinematic studies using RR~Lyrae stars in the solar neighbourhood \citep[e.g.,][]{Layden1996,Maintz2005,Marsakov2018} and in the Galactic bulge \citep{Kunder2016,Kunder2019,Prudil2019Kin}, suggest that a small fraction of nearby RR~Lyrae variables might be associated with the Galactic disc and the old spheroidal component of the Galactic bulge whereas most of them belong to the halo. Mainly the work by \cite{Layden1996} and \cite{Marsakov2018,Marsakov2019s} link some of the local RR~Lyrae stars to the Galactic disc based on their kinematical and chemical properties.

In this paper, we study the kinematic distribution of fundamental mode RR~Lyrae stars in the solar neighbourhood based on radial velocities and astrometric parameters (proper motions and parallaxes) from the {\it Gaia} space telescope \citep{Gaia2016,GaiaBrown2018,Lindegren2018}. We report the discovery of a kinematical feature associated with the disc, which is especially apparent in the pericenter distribution of the local RR Lyrae stars. In Sec.~\ref{sec:DataSample} we describe the features of the collected sample. Section \ref{sec:Orbits} discusses calculated orbital properties for individual objects, which show an apparent peak at a pericenter distance of $\sim$7\,kpc. In Section~\ref{sec:disc}, we explore the possibility that the RR~Lyrae stars in the pericenter peak originate in the Galactic disc. Section \ref{sec:Conclus} summarizes our results.

\section{Data}  \label{sec:DataSample}

We collected a sample of fundamental-mode RR~Lyrae stars from various sources with spectroscopically determined metallicities, radial velocities, photometry in optical and infrared passbands, and precise astrometric solutions. We utilized data from the following sources applying several selection criteria:

\begin{itemize}
\item The metallicities -- [Fe/H] \citep[on the][scale]{ZinnAWest1984} radial velocities, and mean infrared $W1$ and $K_{s}$-band magnitudes (with their errors) of fundamental mode RR~Lyrae stars were obtained from the catalogue assembled by \cite{Dambis2013}. The spectroscopic properties of this catalogue come mostly from \cite{Layden1994} \citep[for more references see][]{Dambis2013}. The infrared photometric properties in the \cite{Dambis2013} catalog were acquired by the \textit{Wide-field Infrared Survey Explorer} \citep[$WISE$,][]{Wright2010,Cutri2012} and \textit{Two-Micron Sky Survey} \citep[2MASS,][]{Cutri2003,Skrutskie2006}. From this sample, 7 stars were removed because their mean magnitudes in one of the photometric bands were unknown, which resulted in 362 variables for the following crossmatch.  
\item In the next step we crossmatched the sample with the {\it Gaia} DR2 catalog \protect\citep{GaiaBrown2018,Lindegren2018}, in order to obtain proper motions ($\mu_{\alpha*}$ and $\mu_{\delta}$) and parallaxes ($\varpi$). The {\it Gaia} data release 2 (DR2) catalog contains information about the uncertainties $\sigma$ and the correlations $\rho$ between the astrometric parameters. In addition, it includes several statistical parameters that indicate the quality of the astrometric solution, e.g. the re-normalized unit weight error (RUWE), which we used to robustly select stars with reliable proper motions. First, we constructed the covariance matrix $\mathbf{\Sigma}$ for proper motions in right ascension $\mu_{\alpha*}$ and declination $\mu_{\delta}$, which we then scaled by the RUWE factor: 
\begin{equation}
\mathbf{\Sigma} = \begin{pmatrix}
\sigma_{\mu_{\alpha*}}^2 & \rho_{\mu_{\alpha*}\mu_{\delta}} \sigma_{\mu_{\alpha*}} \sigma_{\mu_{\delta}}\\ 
\rho_{\mu_{\alpha*}\mu_{\delta}} \sigma_{\mu_{\alpha*}} \sigma_{\mu_{\delta}} & \sigma_{\mu_{\delta}}^2
\end{pmatrix} \cdot \text{RUWE}^{2}
\end{equation} 
In order to introduce a sensible cut on the proper motions based on their errors we needed to diagonalize the covariance matrix by determining its eigenvectors and composing them in the nonsingular matrix $\mathbf{S}$:
\begin{equation}
\mathbf{D} = \mathbf{S}^{-1} \cdot \mathbf{\Sigma} \cdot \mathbf{S},
\end{equation}
where the diagonal matrix $\mathbf{D}$ contains the eigenvalues of $\Sigma$. Using the matrix $\mathbf{S}$ we transformed the vector $\mathbfit{v} = (\mu_{\alpha*}, \mu_{\delta})$ containing the stars' proper motions:
\begin{equation}
\mathbf{V} = ( \mathbf{S}^{-1} \cdot \mathbfit{v} )^{2}.
\end{equation}
Then for the diagonalized covariance matrix $\mathbf{D}$ and the transformed proper motion vector $\mathbf{V}$ we demanded at least 5$\sigma$ significance of the transformed proper motions:
\begin{equation}
\sqrt{ \sum_{k=1}^{2} \mathbf{V} / \text{diag}(\mathbf{D}) } > 5.
\end{equation}
In addition to the proper motion criterion, we also require the same significance for the {\it Gaia} DR2 parallaxes
\begin{equation} 
\varpi/\sigma_{\varpi} > 5.
\end{equation}
Applying these criteria, we removed over 39 stars from our sample. According to \cite{Lindegren2018} the {\it Gaia} DR2 parallaxes are offset from the parallaxes of quasars by $-0.029$\,mas, so we applied this offset to our sample of RR~Lyrae stars. Once the DR2 parallax offset is properly corrected for, the use of Gaia DR2 parallaxes to establish the distances of the local RRLs appears to be effective \citep{Muraveva2018}.
\item The photometric light curves in the $V$-band were collected from the All-Sky Automated Survey for Supernovae \protect\citep[ASAS-SN,][]{Shappee2014,Kochanek2017,Jayasinghe2018}. 
\end{itemize}

In addition to the aforementioned criteria, we removed the stars BI Tel, VX Ind, V363 Cas, V338 Pup, and SS Gru due to their uncertain classification as RR~Lyrae stars \citep[see The International Variable Star Index, VSX, about these objects\footnote{\url{https://www.aavso.org/vsx/index.php}},][]{Watson2006VSX}. Moreover, Y Oct, VY Lib, and MS Ara were removed due to either insufficient data in ASAS-SN or due to their classification as double-mode RR~Lyrae pulsators.  

The mean $V$-band magnitude of an RR Lyrae star was computed as the intercept of a truncated Fourier series fitted to the ASAS-SN photometric data with the following form:
\begin{equation} \label{eq:FourierSeries}
m\left ( t \right ) = A_{0}^{V} + \sum_{k=1}^{n} A_{k}^{V} \cdot \text{cos} \left (2 \upi k \vartheta + \varphi_{k}^{V} \right ),
\end{equation}
where $A_{k}^{I}$ represents the amplitudes, $n$ stands for the degree of the Fourier series, and $\varphi_{k}^{V}$ stands for the phases. The variable $\vartheta$ describes the phase function, which is defined as $\left(HJD-M_{0}\right)/P$, where the $HJD$ represents the time of observation in the Heliocentric Julian Date, and $M_{0}$ and $P$ are the ephemerids, i.e. the time of maximum brightness and the pulsation period. Using the ASAS-SN photometry we also estimated pulsation periods for all studied variables using the Lomb-Scargle periodogram \citep{Lomb1976}. For the errors on the mean $V$-band magnitudes we assumed average photometric errors for the individual variables of $\approx 0.046$\,mag. 

We also computed amplitude ratios and generalized phase differences ($R_{21}$, $\varphi_{21}$, $R_{31}$, $\varphi_{31}$) from the Fourier decompositions:
\begin{equation} \label{eq:FourCoef}
R_{i1} = \frac{A_{i}}{A_{1}}  \\
\varphi_{i1} = \varphi_{i} - i\varphi_{1}
\end{equation}
The amplitude ratios $R_{21}$ and $R_{31}$ reflect the light curve deviation from a sinusoidal shape. With higher values of $R_{21}$ and $R_{31}$ the light curve is more skewed and asymmetric. The phase differences $\varphi_{21}$ and $\varphi_{31}$ mirror the width of the light curve. With decreasing phase differences the light curves become more acute (narrow) at the half maximum light \citep{Simon1988}. The Fourier parameters are useful for the classification of variable stars and for separating them into subclasses. Using the classification from ASAS-SN and our analysis we removed stars that could not be classified as fundamental mode RR~Lyrae stars (for more details see Sec.~\ref{subsec:Missclass} where we discuss the photometric properties of the studied RR~Lyrae variables). 

For each individual star in our sample we determined the extinction in the $W1$, $K_{\rm s}$, and $V$ band using the reddening maps from \cite{Schlafly2011} with the relations $A_{W1} = 0.065\cdot A_{V}$ \citep{Madore2013}, and $A_{K} = 0.114\cdot A_{V}$ \citep{Cardelli1989}. This allowed us to estimate the absolute magnitudes of our stars and to calculate the dereddened mean observed magnitudes in the aformentioned passbands, using \textit{Gaia} parallaxes (incorporated in the Monte Carlo error analysis, see Sec.~\ref{sec:Orbits}). 

We emphasize that throughout this paper we do not use any period-metallicity-luminosity relations to estimate distances for our studied stars. We base our findings on the distance deduced from {\it Gaia} parallaxes.

\begin{figure}
\includegraphics[width=\columnwidth]{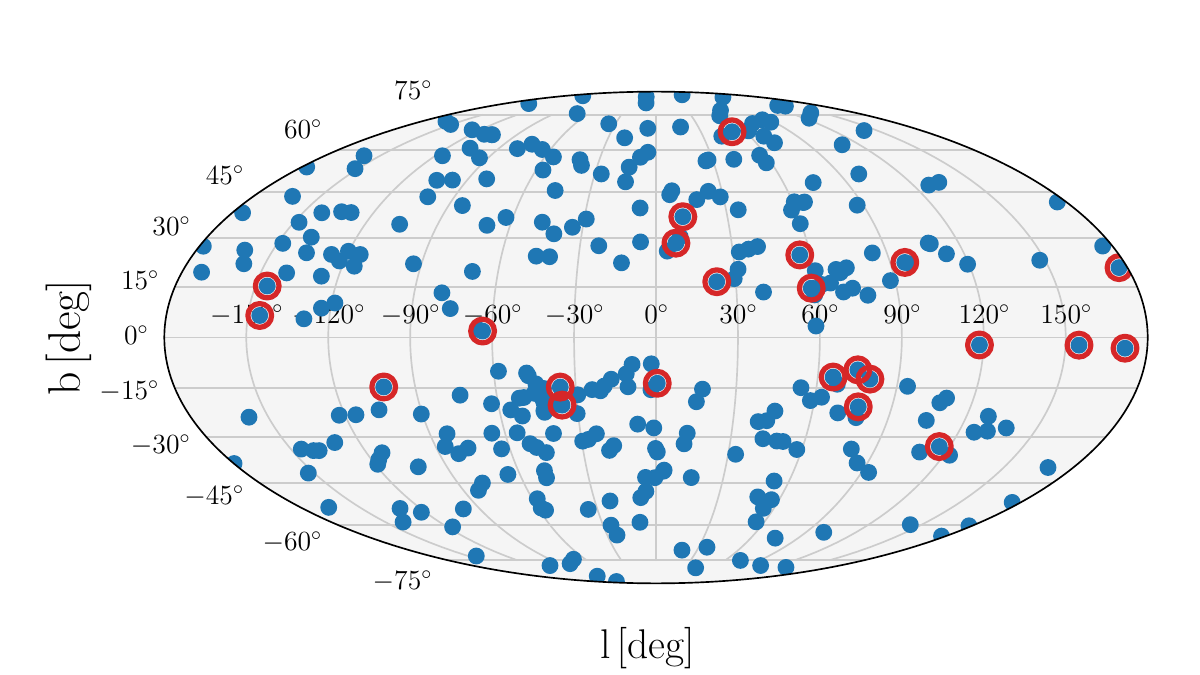}
\caption{The spatial distribution of the studied stars (blue points) in the Galactic coordinates. The red circles encompasing some of the sample RR~Lyrae stars will be explained in the following Section~\ref{sec:Orbits}.}
\label{fig:lbspatial}
\end{figure}

In the end, we had a final sample of 314 RR~Lyrae stars with full kinematic, spatial, and metallicity information, which form the basis of our study. In Fig.~\ref{fig:lbspatial} we depict the spatial distribution of the whole RR~Lyrae sample in Galactic coordinates. In Fig.~\ref{fig:lbspatial} we see that our sample covers the majority of the sky, with the exception of the Galactic bulge and some parts of the Galactic disc. The heliocentric distances of our RR~Lyrae stars range from 0.5\,kpc up to 7\,kpc (with over 90\,\% residing at distances $<$ 3.5\,kpc). The stars cover a broad range in metallicities of $-2.84 < \text{[Fe/H]} < 0.07$\,dex. In apparent magnitude space, almost 97\,\% of the RR~Lyrae variables in our sample are brighter than 14\,mag in the $V$ passband. For comparison, the ASAS-SN survey contains 2778 fundamental mode RR~Lyrae stars brighter than 14\,mag covering a similar coordinate space as in our dataset. Therefore, based on the ASAS-SN catalog, our sample contains approximately 10\,\% of the local RR~Lyrae stars. We emphasize that our sample includes an obvious selection effect, namely the lack of stars toward very low Galactic latitudes, due to the strongly reduced completeness of the ASAS-SN catalog toward these highly attenuated regions. However, we would expect to find more RR~Lyrae stars with thin disc kinematics if our sample was complete toward the Galactic plane. Toward other sight-lines, our selection function does not have any significant non-uniformity.

\section{Orbits} \label{sec:Orbits}

We used the \texttt{galpy}\footnote{Available at \url{http://github.com/jobovy/galpy}} library \citep{galpy2015} to study the kinematics of our sample. For the calculation of the orbital parameters we used the \texttt{MWPotential2014} provided in \texttt{galpy} as an axisymmetric Galactic potential for the MW. The \texttt{MWPotential2014} is composed of three potentials for the Galactic bulge (implemented as a power-law density profile for spherical bulges with an exponential cut-off), a Navarro-Frenk-White halo potential \citep{Navarro1997}, and a Miyamoto-Nagai disc \citep{MiyamotoNagai1975}. For the Sun's position in the MW we assumed a distance to the Galactic center of $R_{0} = 8.178$\,kpc \citep{DistantoSMBH2019} and a height above the Galactic plane of $z_{\odot } = 25$\,pc \citep{Juric2008}. In order to correct for solar motion, we used the Sun's velocity with respect to the local standard of rest $\left(U_{\odot},V_{\odot},W_{\odot} \right) = \left(-11.1, 12.24, 7.25\right)$\,km\,s$^{-1}$ \citep{Schonrich2010} and the velocity of the local standard of rest (LSR) $v_{LSR}=235$\,km\,s$^{-1}$ based on $R_{0}$ \citep{DistantoSMBH2019} $R_{0}$, the proper motion of Sgr A* \citep{Reid2004}, and $V_{\odot}$. 

For each star, we integrated over a 1\,Gyr timespan and calculated its orbital parameters: maximum height from the Galactic plane $z_{\text{max}}$, eccentricity $e$, and peri- and apocenters of its orbit, $r_{\text{per}}$ and $r_{\text{apo}}$, respectively (a 100\,Myr, 300\,Myr or 500\,Myr timespan does not affect the orbital parameters significantly). Using \texttt{galpy}, we calculated the full 6D solution (spatial and kinematical) for the studied sample. With the \texttt{MWPotential2014} we also calculated the Hamiltonian action integrals of motion in an axisymmetric potential $\mathbf{J} = \left( J_{R}, J_{\phi}, J_{z}\right) $, where $J_{R}$ represents the oscillation in the radial direction, $J_{\phi}$ is the azimuthal action (angular momentum in $z$-direction $L_{z}$), and $J_{z}$ describes the vertical oscillation.

A Monte Carlo error analysis was performed for the complete calculation taking into account the covariance matrix of the {\it Gaia} astrometric solution, errors in radial velocities, reddening, and mean magnitudes. The resulting values and their errors were taken as the median, first, and third quartile from the generated distributions, in addition, we also obtained correlations between orbital parameters and velocities. The first few lines of a table with calculated photometric, chemical and orbital properties can be found in the Appendix of this paper\footnote{The full table can be found in the supplementary material}.  

In the top panel of Fig.~\ref{fig:OrbitParam} we show a histogram for the distribution of the pericentric distances in our sample. In this distribution, we noticed a small peak approximately at $r_{\text{per}} = 7$\,kpc (marked with an arrow). To verify its existence we generated 1000 random realizations of the pericentric distribution for the studied stars using their median values and quantiles. For each generated distribution we calculated the \texttt{Gaussian mixture model} probability distribution implemented in the \texttt{scikit-learn} library \citep{Pedregosa2011}. Using the Bayesian information criterion (BIC) and the Akaike information criterion (AIC), we estimated a suitable number of Gaussian components for each distribution. For the BIC, in nearly all cases (above 90\,\%) the suggested number of Gaussians was 3, while for the AIC, three components were suitable in more than 42\,\% of the cases. In all of the pericentric variations, the peak is present in the histogram and is modeled by the Gaussian mixture model. The generated distributions were used to estimate the errors of the pericentric distribution. Using the calculated errors, the probability density of the pericentric peak ($0.0813 \pm 0.0065$) has a 3$\sigma$ significance in comparison with the neighboring valley ($0.0460 \pm 0.0092$, marked with black arrows in Fig.~\ref{fig:OrbitParam}). 

It is worth noting that this overdensity in pericenter distance is also present when using the older pericenter values calculated by \cite{Maintz2005}, but was not specifically pointed out in their or any previous studies of the local RR~Lyrae stars. We also observe a peak around 3\,kpc, which is modeled by one of the Gaussian components. We tentatively associate this peak with RR~Lyrae stars in the Galactic thick disc.

For comparison with our sample, we calculated the orbital solution for the stars observed in the Radial Velocity Experiment (RAVE) data release 5 \citep{Kunder2017}. RAVE is a spectroscopic multi-fiber survey of Milky Way stars in the $9 < I < 12$ magnitude range \citep{Steinmetz2006}. Because of its magnitude range, RAVE observed mainly the thick and thin disc stars \citep[see fig.~19 in][]{Kordopatis2013}. We crossmatched the RAVE sample with {\it Gaia} DR2, requiring at least 5$\sigma$ significance for proper motions and parallaxes. The distribution of pericentric distances in the RAVE sample is shown in the top panel of Fig.~\ref{fig:OrbitParam} in grey colour. From these two distributions, we see that RR~Lyrae stars at $r_{\text{per}} \approx 7$\,kpc fall into the region where the majority of the RAVE disc sample lies.

This small overdensity is visible also in other orbital parameters, e.g., in $e$ and $z_{\text{max}}$ (see the bottom panel of Fig.~\ref{fig:OrbitParam}), where the majority of stars with $r_{\text{per}} \approx$ 7\,kpc clump at low values of eccentricity ($e<0.4$) with rather small excursions above the Galactic plane ($z_{\text{max}}<2$\,kpc) is located. The low values of $e$ together with marginal $z_{\text{max}}$ suggest rather circular orbits confined to the Galactic disc. Including another dimension in the form of metallicity in the bottom panel of Fig.~\ref{fig:OrbitParam} reveals that the majority of the most metal-rich RR~Lyrae stars from our sample show low eccentricities and $z_{\text{max}}$ around the pericentric distance of this peak. We note that RR~Lyrae stars with pericentric distances between 5-9\,kpc and eccentricity above 0.5 contribute partially to the pericenter peak shown above, however, the peak is still significant after their removal.

To further isolate RR~Lyrae stars sharing the same kinematics and possibly contributing to the pericenter peak, we used two criteria:
\begin{equation} \label{eq:conditions} 
e < 0.2 \\
z_{\text{max}} < 0.9\,\text{kpc}
\end{equation}
The additional condition on eccentricity $e$ narrows the sample to circular orbits common for stars in the Galactic disc \citep{Hayden2019}\footnote{More than 70\,\% of stars from their sample in the solar neighbourhood ($d < 500$\,pc) have an eccentricity $e<0.2$.}. The criterion on the excursion from the Galactic plane $z_{\text{max}}$ is consistent with the MW thick disc scale height $H \approx 0.9$\,kpc \citep{Juric2008,BlandHawthorn2016}. We did not employ any criteria for pericentric and apocentric distances. These two conditions yield 22 possible disc-like RR~Lyrae stars, which are denoted by red circles in the Fig.~\ref{fig:lbspatial} (also marked with asterisks in Tab.~\ref{tab:KinProp}). We note that almost 90\,\% of the stars in the RAVE sample fulfill these conditions as well.

\begin{figure}
\includegraphics[width=\columnwidth]{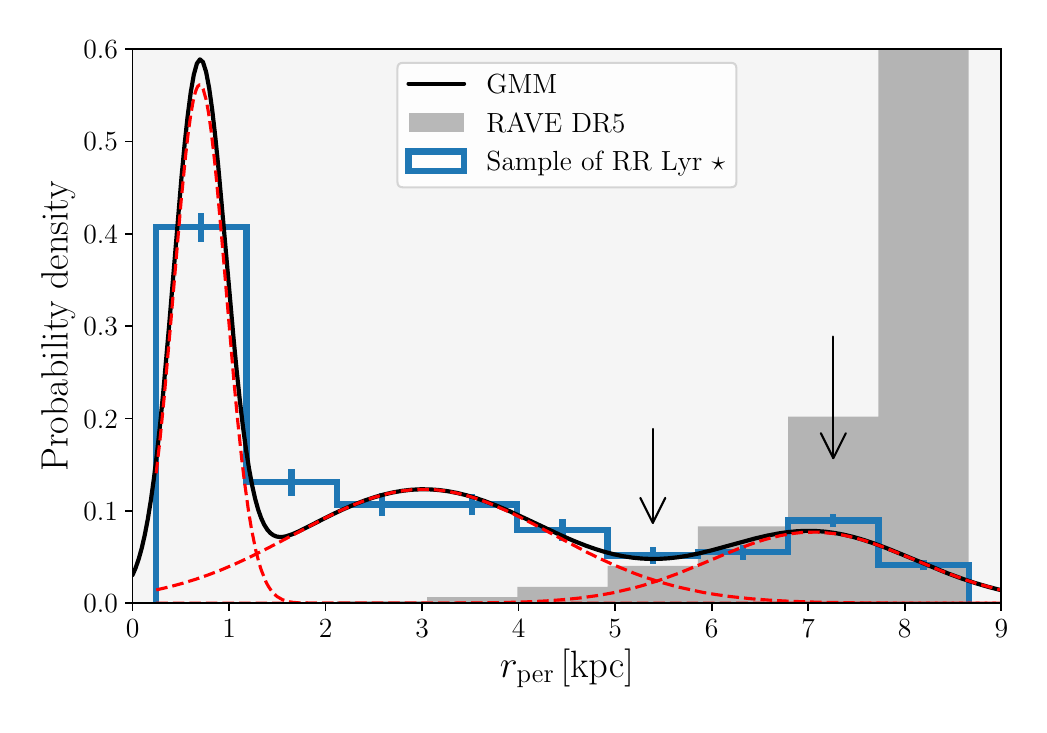}
\includegraphics[width=\columnwidth]{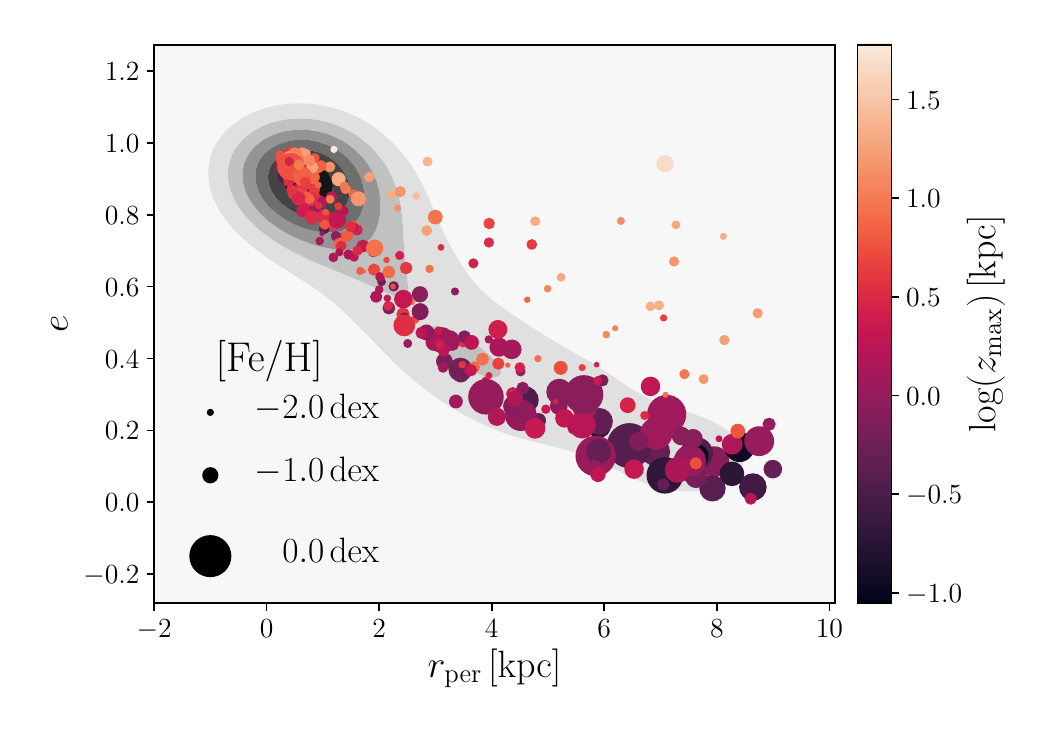}
\caption{Distribution of pericentric distances $r_{\text{per}}$ (top panel) and eccentricity $e$ vs. $r_{\text{per}}$ dependence (bottom panel) for our sample. In the top panel, the black solid line depicts the \texttt{Gaussian mixture model} with individual components denoted by red dashed lines. The grey distribution represents stars from the RAVE survey, while the blue columns stand for our sample of RR~Lyrae variables. The black arrow points toward the pericentric peak of the local RR~Lyrae stars. The error bars on individual columns were calculated from the Monte Carlo simulation with a fixed bin size. The bottom panel shows $e$ vs. $r_{\text{per}}$ with a colour-coding based on $z_{\text{max}}$. The sizes of each point indicate the metallicity of individual RR~Lyrae stars. The grey background shading represents the kernel density estimates (KDE) of the overlay points.}
\label{fig:OrbitParam}
\end{figure}

\begin{figure}
\includegraphics[width=\columnwidth]{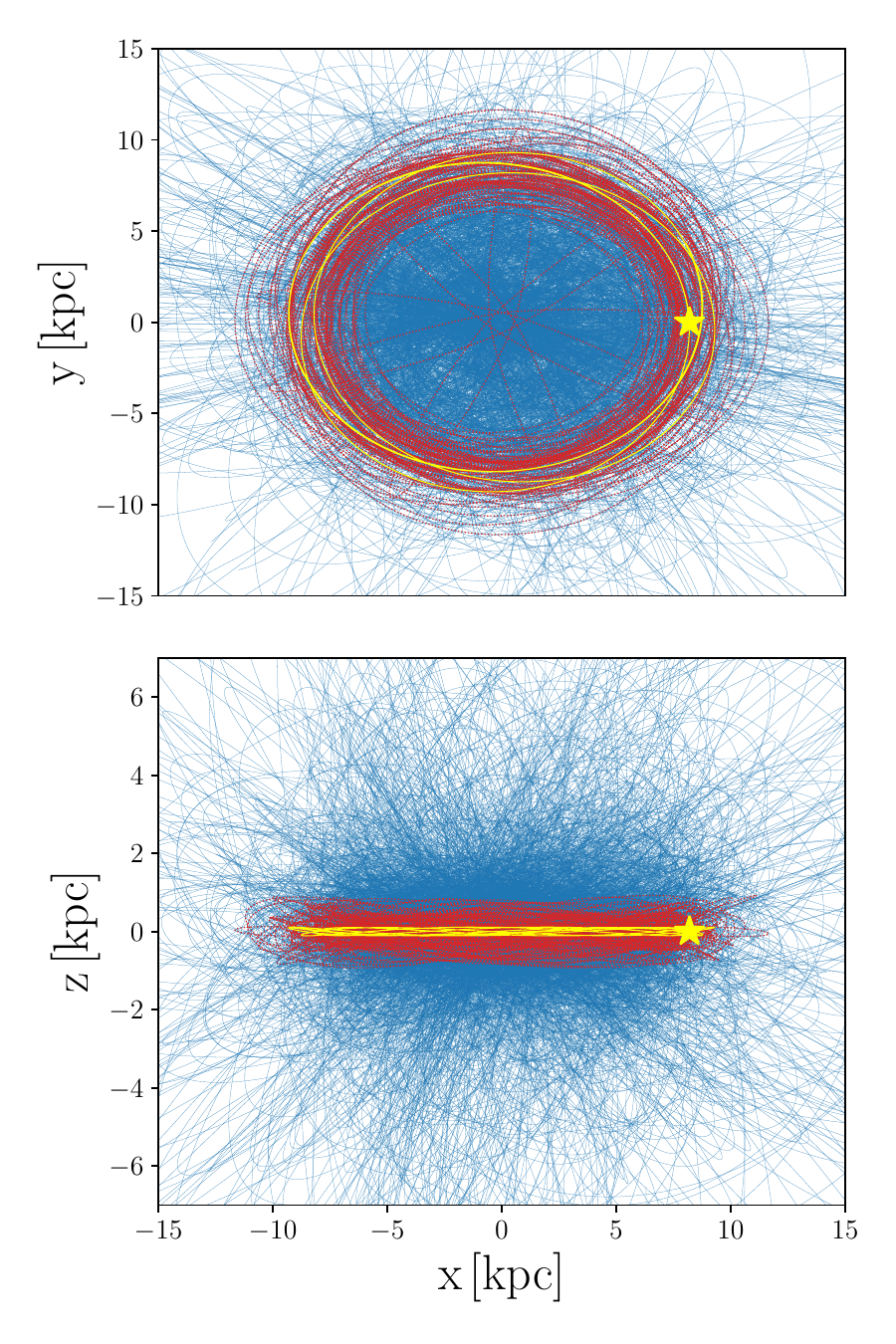}
\caption{The spatial distrbution of the studied stars in Cartesian rectangular Galactocentric coordinates, with the top panel depicting the $y$ vs. $x$ and the bottom panel showing the $z$ vs. $x$ distributions, respectively. The blue lines stand for orbits of the sample stars, while the red dotted lines represent RR~Lyrae stars based on the conditions in Eq.~\ref{eq:conditions}. The yellow stars and lines mark the position and orbit of the Sun, respectively.}
\label{fig:XYCoordinates}
\end{figure}

The full orbital solution using \texttt{MWPotential2014} is examined in Fig.~\ref{fig:XYCoordinates}, where we show the distribution of orbits of the whole sample (blue solid lines) in rectangular Galactocentric coordinates. RR~Lyrae variables fulfilling Eq.~\ref{eq:conditions} (red dotted lines) are concentrated at small heights above the plane with almost circular orbits. This implies their association with the Galactic disc, which will be examined in the following chapters. We emphasize that throughout the paper, we refer to these 22 RR~Lyrae stars as disc stars, as their association with either the thin or thick disc is not clear when based on orbital parameters alone. In Section~\ref{sec:disc}, their association with these disc components is explored. The conditions in Eq.~\ref{eq:conditions} lead to a sample of disc-like RR~Lyrae stars with almost identical orbits (see Fig.~\ref{fig:XYCoordinates}). With regard to the apocentric distances $r_{\text{apo}}$, the selected disc RR~Lyrae stars vary between $\approx$ $7.9 - 11.5$\,kpc, which is expected due to our condition on the eccentricity of the orbits, where eccentricity is connected with $r_{\text{apo}}$ through following relation:
\begin{equation}
e = \frac{r_{\text{apo}} - r_{\text{per}}}{r_{\text{apo}} + r_{\text{per}}}.
\end{equation} 
The 22 RR~Lyrae stars selected based on Eq.~\ref{eq:conditions} will be examined for their possible disc association in the following Section.

\section{RR~Lyrae stars in the Galactic disc} \label{sec:disc}

The idea that some of the RR~Lyrae stars in the solar neighbourhood originated in the Galactic disc is not new. \cite{Layden1996} suggested that some field RR~Lyrae stars belong to the Galactic thick disc, as did \cite{Marsakov2018}. \cite{Layden1996} suggested several approaches to separate disc and halo RR~Lyrae stars based on their kinematics (see below) and metallicities, suggesting that disc and halo separate at [Fe/H]$ = -1$\,dex for RR~Lyrae stars. Here the chemodynamics of the pericenter peak RR~Lyrae are explored.

\subsection{Kinematic tests}

\begin{figure}
\includegraphics[width=\columnwidth]{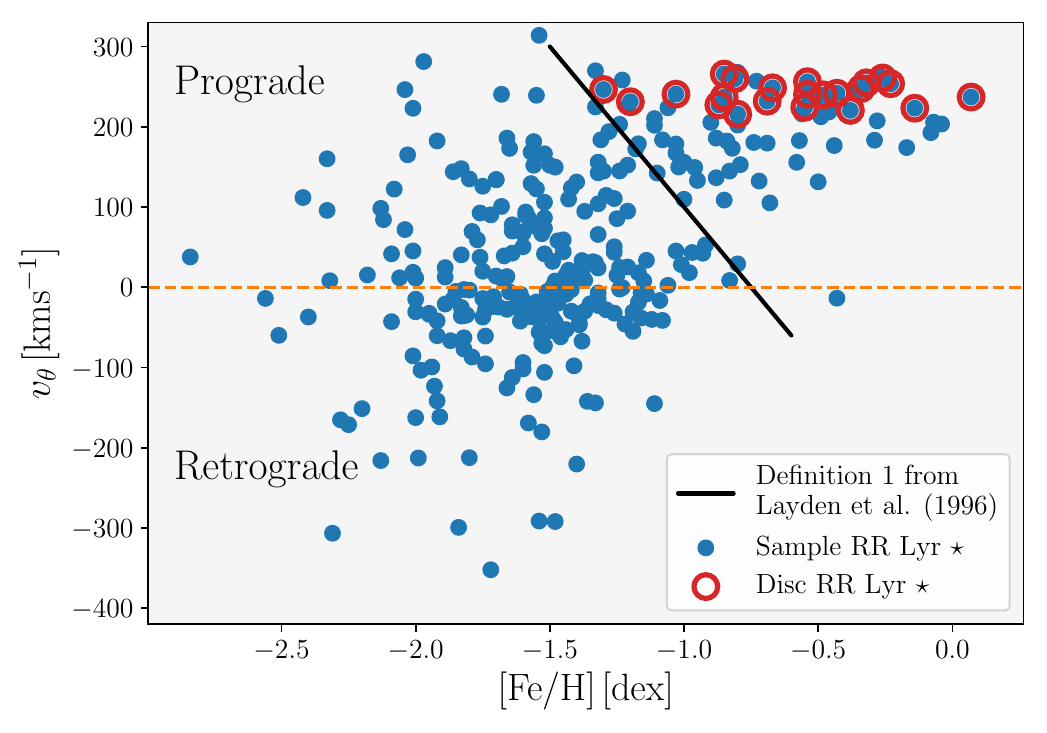}
\caption{The tangential velocity $v_{\theta}$ vs. metallicity [Fe/H] for the studied variables (blue points). The variables associated with the pericenter peak (based on our criteria in Eq.~\ref{eq:conditions}) are highlighted with red circles. The black solid line represents definition 1: $v_{\theta}= -400 \cdot$[Fe/H]$-300$ listed in table 3 of \citet{Layden1996}, and the dashed orange line represents the separation of stars with prograde/retrograde motion based on their $v_{\theta}$.}
\label{fig:FeV}
\end{figure}

Although we are using almost the identical sample as used by \cite{Layden1996}, {\it Gaia} DR2 affords a more precise astrometric solution, and thus better constrained orbits for individual stars. In Fig.~\ref{fig:FeV} we show the tangential velocity $v_{\theta}$ vs. metallicity [Fe/H] distribution, where $v_{\theta}$ represents the Galactocentric cylindrical velocity, positive in the direction of the Galactic rotation. Using definition 1 in table 3 from \cite{Layden1996} (black solid line in Fig.~\ref{fig:FeV}), we separated our star sample into disc and halo RR~Lyrae variables. We see that the stars in the pericenter peak (Eq.~\ref{eq:conditions}) all fall in the {\it disc} region in $v_{\theta}$ vs. [Fe/H] space. We note that other stars identified by \citeauthor{Layden1996} as the Galactic disc RR~Lyrae variables show resemblence to the stars identified using the criteria in Eq.~\ref{eq:conditions}, but they fail our criteria mainly on the basis of $r_{\text{per}}$ and $e$, where approximately half of the stars have $e>0.2$. 

\begin{figure}
\includegraphics[width=\columnwidth]{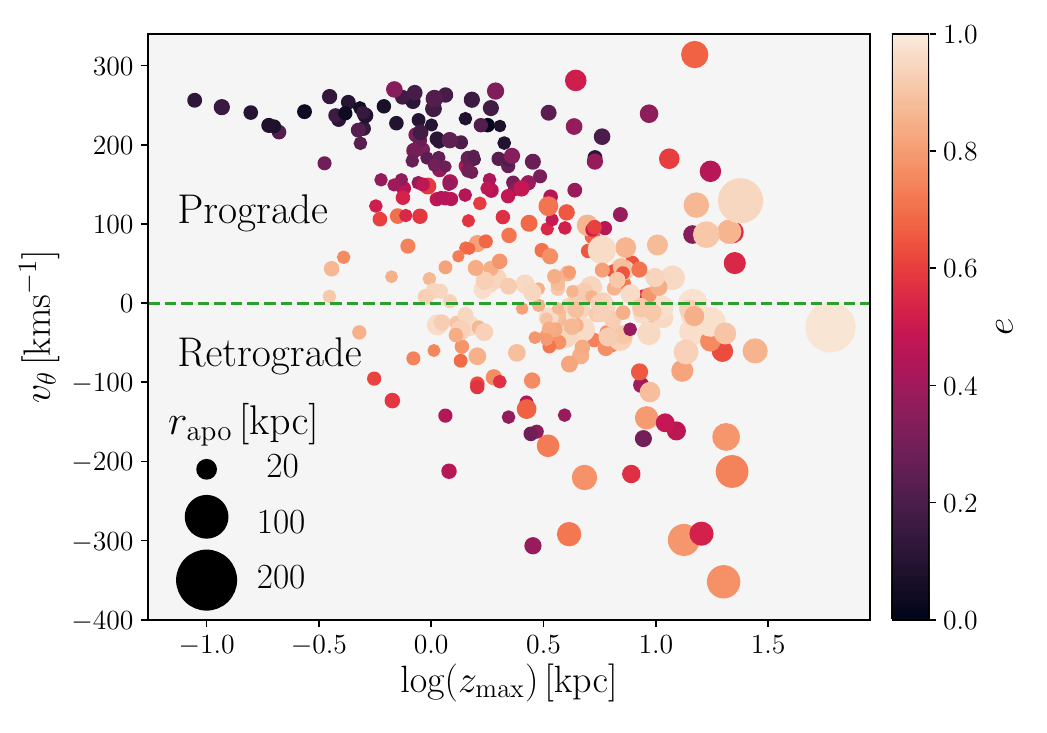}
\caption{The tangential velocity $v_{\theta}$ vs. maximum height from the Galactic plane $z_{\text{max}}$ for the studied RR~Lyrae stars. The colour-coding of the individual points is based on $e$, and size of each point coresponds to the $r_{\text{apo}}$.}
\label{fig:Apo}
\end{figure}

We note that $\approx 39\,\%$ of the studied variables have retrograde rotation ($v_{\theta}$ < 0\,km\,s$^{-1}$) and nearly all stars on retrograde orbits have metallicities of [Fe/H]$<-1$\,dex. We can crudely distinguish RR~Lyrae stars formed in-situ from accreted stars based on their prograde/retrograde motion in the Galaxy. In our RR~Lyrae dataset, retrogradely moving variables exhibit on average higher values for $r_{\text{apo}}$, $e$, and $z_{\text{max}}$ in comparison with the rest of the variables (see Fig.~\ref{fig:Apo}). This is probably due to the presence of disc RR~Lyrae stars in the prograde part of our sample. Together with their on average lower $r_{\text{per}}$ (with respect to the remaining pulsators) this supports their possible extragalactic origin \citep{Villalobos2009,Qu2011}.

\subsubsection{Velocity components}

\begin{figure}
\includegraphics[width=\columnwidth]{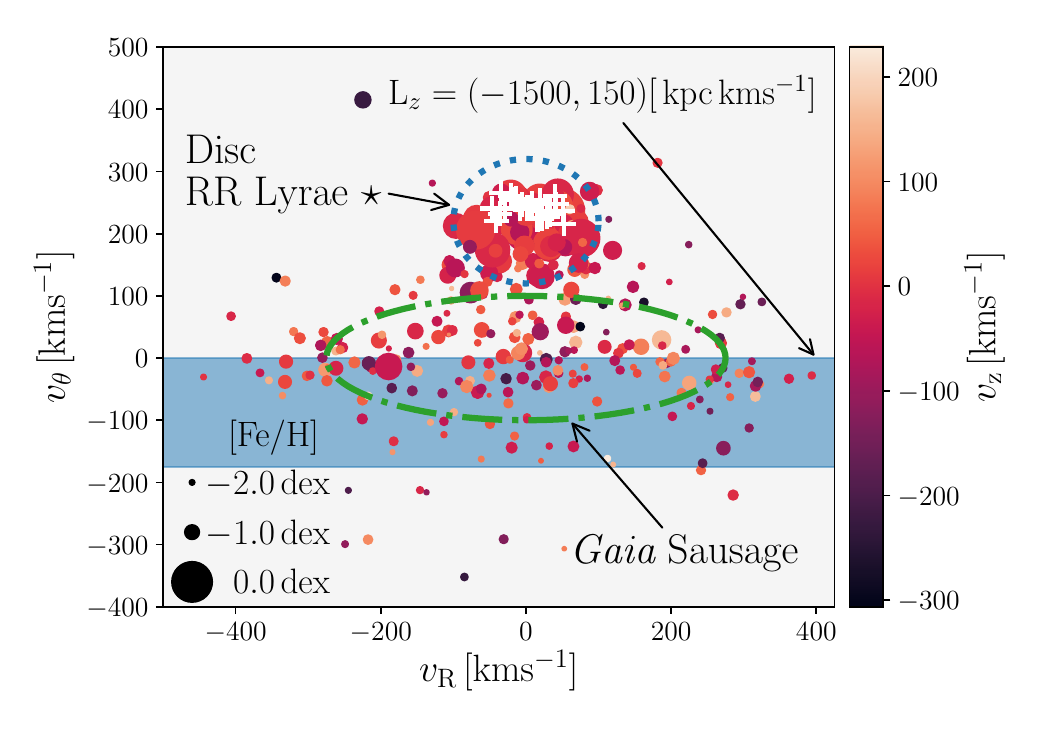}
\caption{The distribution of the velocity components $v_{\theta}$ vs. $v_{\rm R}$ for the studied RR~Lyrae stars. The colour-coding of the individual points is based on their transversal velocity $v_{\rm z}$. The point sizes reflect their metallicity with decreasing size for decreasing metallicity. The white crosses mark the positions of pericenter peak RR~Lyrae variables. The area of their occurrence is also highlighted with a blue dotted line. The green dashed line represents the approximate region for stars associated with the {\it Gaia}-Sausage in $v_{\theta}$ vs. $v_{\rm R}$ space based on \citet{Belokurov2018}. In addition, the blue stripe covers the region where the vast majority of our sample RR~Lyrae stars fulfill the condition $-1500$\,kpc\,km\,s$^{-1}<L_{z}<150$\,kpc\,km\,s$^{-1}$ \protect\citep[from ][]{Helmi2018Nature}, as discussed below.}
\label{fig:SausageVtheta}
\end{figure}

The pericenter peak RR~Lyrae variables are also distinguishable in the distribution of their velocity components (see Fig.~\ref{fig:SausageVtheta}). Here disc stars (roughly marked with a blue ellipse) would reside in the top part of the $v_{\theta}$ vs. $v_{\rm R}$ distribution. The average values and dispersion for the depicted disc RR~Lyrae stars in Fig.~\ref{fig:SausageVtheta} are $v_{\rm R} = 4 \pm 30$\,kms$^{-1}$ and $v_{\theta} = 241 \pm 14$\,kms$^{-1}$. From Fig.~\ref{fig:SausageVtheta} we also see that the transversal velocity $v_{\rm z}$ for disc stars varies around 0\,km\,s$^{-1}$, which is expected based on our cut in the $z_{\text{max}}$.

The {\it Gaia}-Enceladus or {\it Gaia} Sausage \citep{Belokurov2018,Helmi2018Nature,Gallart2019}, is clearly seen among the local RR Lyrae stars. This remnant of a massive merger of a Small Magellanic Cloud-like galaxy in the local stellar halo contains rather metal-poor RR~Lyrae stars ([Fe/H]$>-1.5$\,dex) with a highly non-Gaussian velocity ellipsoid. There have been several studies linking some of the MW stars and globular clusters with {\it Gaia}-Enceladus, e.g. \cite{Helmi2018Nature,Myeong2018}. These studies have an approximately common range of angular momenta in $z$-direction, $L_{z}$, with values $-1500$\,kpc\,km\,s$^{-1}<L_{z}<150$\,kpc\,km\,s$^{-1}$. The majority of the stars with values of $L_{z}$ within the aforementioned boundaries are located in the blue shaded region of Fig.~\ref{fig:SausageVtheta}. Some of the MW RR~Lyrae stars have already been associated with {\it Gaia}-Enceladus by \cite{Simion2019} while studying the Hercules-Aquila Cloud and Virgo Overdensity \citep{Vivas2001,Newberg2002,Belokurov2007}. This is the first time we can link {\it Gaia}-Enceladus with the local RR~Lyrae stars \cite[based on $L_{z}$ and the region covered by ellipse in $v_{\theta}$ vs. $v_{\rm R}$ defined by][]{Belokurov2018} (variables kinematically associated with the {\it Gaia}-Enceladus/Sausage are marked with a plus sign in Tab.~\ref{tab:KinProp}). 

Furthermore, using the \texttt{galpy} potential for the MW, \texttt{MWPotential2014}, we calculated the asymmetric drift of the RR~Lyrae variables associated with the Galactic disc concerning the circular velocity. We found that the studied subpopulation of the local RR~Lyrae stars lags behind the Galactic rotation with an asymmetric drift equal to $-5$\,kms$^{-1}$, which in agreement with the asymmetric drift assumed for the thin disc stars \citep[see e.g.,][]{Bensby2003,Golubov2013,Sysoliatina2018}. In addition, pericenter peak RR~Lyrae stars appear to be kinematically cold, with a vertical velocity dispersion $\sigma_{v_{\rm z}} = 16$\,kms$^{-1}$ and a total velocity dispersion\footnote{$\sigma_{s} = \sqrt{\sigma_{v_{\theta}}^{2} + \sigma_{v_{\rm R}}^{2} + \sigma_{v_{\rm z}}^{2}}$} equal to $\sigma_{s} = 37$\,kms$^{-1}$, which agrees with their possible thin disc association.

Here we would like to add a note on high-velocity RR~Lyrae stars in our sample. RR~Lyrae variables with high space velocity have been reported in the Galactic bulge \citep[see e.g.,][]{Kunder2015,Hansen2016}. One example of a high-velocity RR~Lyrae star is seen in the top left corner of Fig.~\ref{fig:SausageVtheta} (dark point, $v_{\theta}=415$\,kms$^{-1}$ and $v_{\rm R}=-224$\,kms$^{-1}$), AO Peg, with [Fe/H]$ = - 0.92$\,dex and a space velocity $sv = 512$\,km\,s$^{-1}$. In general, for only 2\,\% of the RR~Lyrae pulsators from our sample the $sv$ exceed $400$\,km\,s$^{-1}$, which is similar to RR~Lyrae stars in the Galactic bulge sample \citep{Prudil2019Kin} where we found roughly 3\,\% of the RR~Lyrae stars to show $sv$ above $400$\,km\,s$^{-1}$.

\subsubsection{Association based on relative probabilities}

A probabilistic approach to separate stars in the MW components (thin and thick disc and halo) was proposed by \cite{Bensby2003}. The selection is done strictly on a kinematic basis using the Galactocentric rectangular velocities relative to the local standard of rest (LSR) -- $U_{\rm LSR}, V_{\rm LSR}, W_{\rm LSR}$, while $U_{\rm LSR}$ in the left-handed system is positive in the Galactic anticenter direction, $V_{\rm LSR}$ increases in direction of the Galactic rotation, and the positive direction toward the Galactic north pole is described by increasing $W_{\rm LSR}$. The method introduced by \cite{Bensby2003} assumes that the aforementioned velocities for stars in the MW components follow Gaussian distributions:
\begin{align} \label{eq:bensby2003}
\begin{split}
f = \frac{1}{\left ( 2\upi \right )^{3/2} \sigma_{U_{\rm LSR}} \sigma_{V_{\rm LSR}} \sigma_{W_{\rm LSR}}} \cdot \\ \text{exp} \left ( -\frac{U_{\rm LSR}^{2}}{2\sigma_{U_{\rm LSR}}^{2}} -\frac{\left ( V_{\rm LSR} - V_{\rm asym} \right )^{2}}{\sigma_{V_{\rm LSR}}^{2}} - \frac{W_{\rm LSR}^{2}}{2\sigma_{W_{\rm LSR}}^{2}} \right ),
\end{split}
\end{align} 
where $V_{\rm asym}$ is the velocity of the asymmetric drift and the $\sigma_{U_{\rm LSR}}$, $\sigma_{V_{\rm LSR}}$, $\sigma_{W_{\rm LSR}}$ represent the velocity dispersions. Our assumed velocity dispersions and velocities of the asymmetric drift for the individual MW components are listed in Tab.~\ref{tab:bensby} \citep{Holmberg2000,Bensby2003}. We emphasize that our sample of variables are Population~II stars with ages above 10\,Gyr, while studies of the MW thin disc propose ages from 0 to 8\,Gyr, e.g., \citep{Haywood2013}. Therefore, based on that age range we do not have a reasonable priors for the probability of RR~Lyrae stars emerging from the thin disc population, since they are believed to be older than 10\,Gyr \citep[e.g.,][]{Glatt2008, VandenBerg2013}. Thus, for this analysis we will assume only two MW components: disc -- D and halo -- H, where for the disc we will assume priors used for the thick disc in \cite{Bensby2003}. For each star from our sample we solve Eq.~\ref{eq:bensby2003} and get probabilities for disc and halo membership: $f_{\rm D}$ and $f_{\rm H}$.

\begin{figure}
\includegraphics[width=\columnwidth]{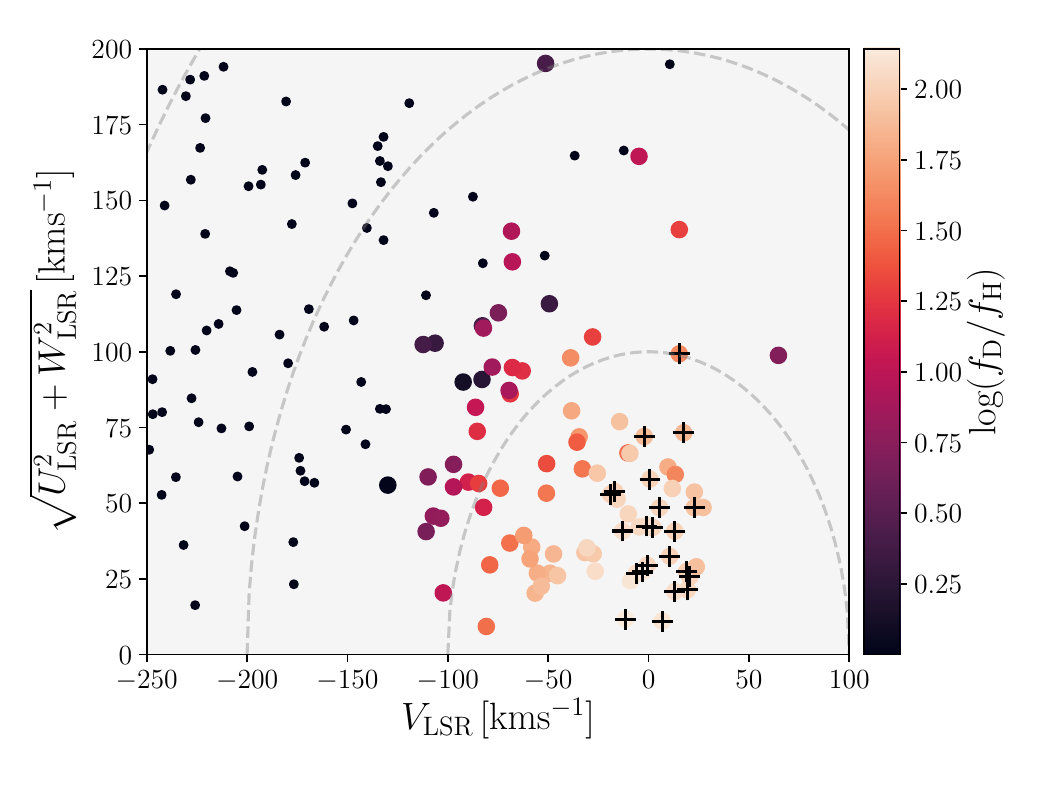}
\caption{Toomre diagram for our studied RR~Lyrae stars constructed using the Galactocentric rectangular velocities. The colour-coding is based on the logarithm of the probability ratio between disc and halo. We note that we removed from the colour-coding stars with $f_{\rm D} /f_{\rm H} < 1$ and denoted them only as black dots since they are not a target of our analysis. The pericenter peak RR~Lyrae stars are marked with black crosses. The same notation applies also for the enlarged inset panel. The grey dashed lines represent lines of constant space velocity with a step size of 50\,km\,s$^{-1}$.}
\label{fig:BensbyToomre}
\end{figure}

The results of this analysis can be seen in the Toomre diagram depicted in Fig.~\ref{fig:BensbyToomre}. Here, our sample of RR~Lyrae stars is colour-coded based on the logarithm of the disc to halo probability ratio log$(f_{\rm D} / f_{\rm H})$. We note that $f_{\rm D} /f_{\rm H} < 1$ for over two-thirds of the sample, thus for clarity of Fig.~\ref{fig:BensbyToomre}, we included stars with $f_{\rm D} /f_{\rm H} < 1$ only as black dots in the aformentioned figure. In the same figure, we see that stars with a high probability ratio ($f_{\rm D} / f_{\rm H}>10$) are concentrated around $V_{\rm LSR} = 0$ where one would expect to find the disc population. All of the disc RR~Lyrae stars have a high disc to halo probability ratio (all above $f_{\rm D} /f_{\rm H} > 42$), which further corroborates their association with the MW disc population. One could argue that we did not use the observed fraction of stars in the solar neighbourhood as in \cite{Bensby2003} for our probability ratio. As a sanity check, we tested various values for the observed fraction of disc and halo stars in the solar neighbourhood, e.g., from \cite{Bensby2003} and \cite{Juric2008}. In the end, the pericenter peak RR~Lyrae variables still had the highest probability ratios among the stars in our sample. 

\begin{table}
\caption{The adopted velocity dispersions and asymmetric drifts based on \protect\cite{Bensby2003}. Column 1 lists two MW components (halo -- H, thick disc -- TD) and column 2 contains the assumed velocities for the asymmetric drift. Columns 3, 4, and 5 provide values for the velocity dispersions for the thick disc and halo.}
\label{tab:bensby}
\begin{tabular}{llllll}
\hline
   MW      & $V_{\rm asym}$ & $\sigma_{U_{\rm LSR}}$ & $\sigma_{V_{\rm LSR}}$ & $\sigma_{W_{\rm LSR}}$\\ 
   & [km\,s$^{-1}$] & [km\,s$^{-1}$] & [km\,s$^{-1}$] & [km\,s$^{-1}$] &       \\ \hline
H      & $-220$            & 160                    & 90                     & 90  \\
TD & $-46$            & 67                     & 38                     & 35    \\ \hline
\end{tabular}
\end{table}

The majority of the local RR Lyrae stars are associated with the Galactic halo \citep{Maintz2005,Mateu2012,Mateu2018,Marsakov2018}, but the stars in the pericenter peak are consistent with the Galactic disc. 

\subsection{Chemical test}

\begin{figure}
\includegraphics[width=\columnwidth]{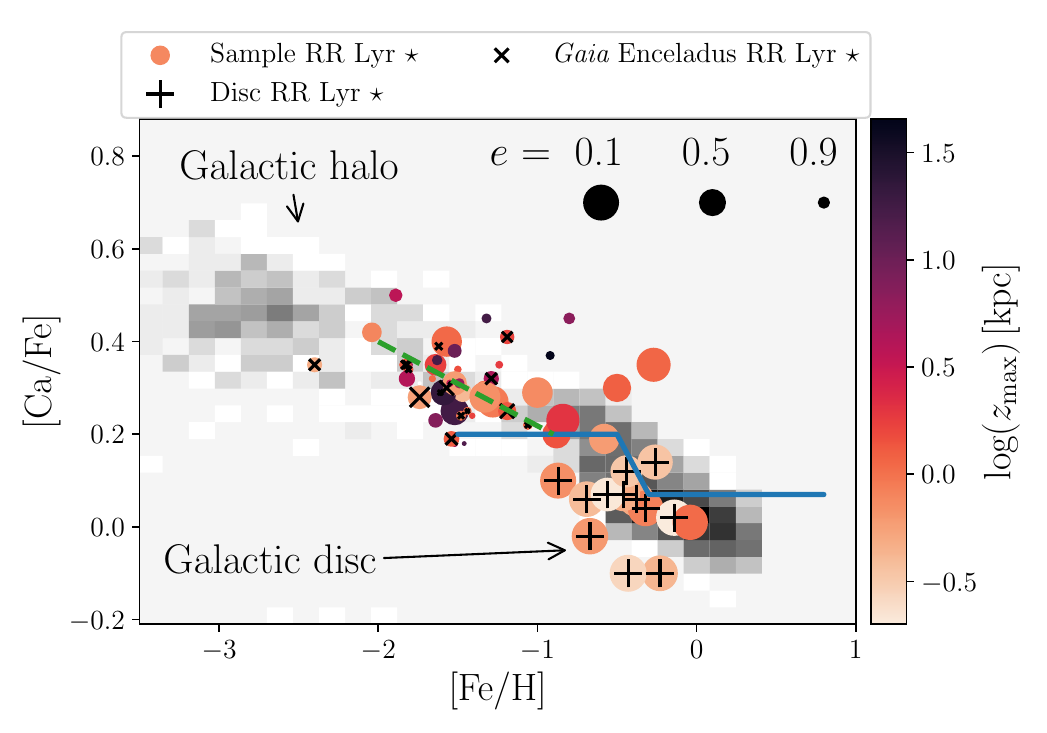}
\caption{The [Ca/Fe] vs. [Fe/H] dependence for the studied RR~Lyrae stars with determined calcium abundance from the literature, colour-coded based on the excursion from the Galactic plane $z_{\text{max}}$, and marker sizes based on the orbital eccentricity $e$. RR~Lyrae stars associated with the pericenter peak based on our set of conditions are marked with a black plus sign. The RR~Lyrae variables possibly associated with the {\it Gaia}-Enceladus ($-1500$\,kpc\,km\,s$^{-1}<L_{z}<150$\,kpc\,km\,s$^{-1}$ and falling in the green ellipse in Fig.~\ref{fig:SausageVtheta}) are marked with black crosses. The green dashed line stands for the boundary between $\alpha$-rich and $\alpha$-poor populations from \citet{Hayes2018}. The blue solid line represents the approximate separation of the thin and thick disc stars \citep{Adibekyan2011}. The underlying gray distribution comes from the spectroscopic studies of non-variable stars in the MW disc and halo.}
\label{fig:ChemicalRRLYR}
\end{figure}

Here we use chemical information together with kinematics to explore the pericenter peak RR~Lyrae stars. A search of the literature was carried out to collect information on the elemental abundances of the pericenter peak RR~Lyare stars. We found 61 RR~Lyrae stars with determined [Ca/Fe] in our sample from the following studies: \cite{Chadid2017,Liu2013,Pancino2015,Lambert1996}. Calcium is an element resulting from helium burning via $\alpha$ particle capture and is synthesized in massive stars. Therefore, it can serve as a proxy for [$\alpha$/Fe]. Fig.~\ref{fig:ChemicalRRLYR} shows [Ca/Fe] vs. [Fe/H] for the collected data overplotted over the distribution of non-variable MW stars \citep{Edvardsson1993,Reddy2003,Bensby2003,Roederer2014,Bensby2014}.

In Fig.~\ref{fig:ChemicalRRLYR} we combined the chemical abundance information with the kinematic properties: The point sizes vary with $e$ and the colour-coding is based on $z_{\text{max}}$. Stars in the pericenter peak defined by the conditions of Eq.~\ref{eq:conditions} are located at low values of [Ca/Fe] typical for thin disc stars \citep[in other words, consistent with the $\alpha$-poor disc,][]{Adibekyan2011,Adibekyan2013,Bland-Hawthorn2019}\footnote{We note that the blue line separating the thick and thin disc was derived for [$\alpha$/Fe] while in our case we have only [Ca/Fe]. The overall difference is negligible.}. The remaining stars are located at the $\alpha$ element knee and further close to the region typical for thick disc stars \citep[the $\alpha$-rich disc,][]{Bland-Hawthorn2019}, and Galactic halo stars. It is striking that the pericenter peak RR~Lyrae stars reside in a different [Ca/Fe] - [Fe/H] space from the rest of the local RR~Lyrae stars.

In addition, the majority of the RR~Lyrae variables (14 out of 17) associated with {\it Gaia}-Enceladus based on $L_{z}$, $v_{\theta}$, and $v_{\rm R}$ with published [Ca/Fe] abundances seems to fall beyond the \textit{knee}, on $\alpha$-poor sequence discovered by \cite{Nissen2010}. This is in agreement with the studies of {\it Gaia}-Enceladus where stars associated with {\it Gaia}-Enceladus exhibit low $\alpha$ abundances at low metallicities \citep{Hayes2018,Haywood2018,Helmi2018Nature}. The RR~Lyrae stars associated with {\it Gaia}-Enceladus also exhibit a considerable metallicity spread in the old population (more than 1\,dex)

\section{Possibility of misclassification} \label{subsec:Missclass}

\begin{figure*}
\includegraphics[width=2\columnwidth]{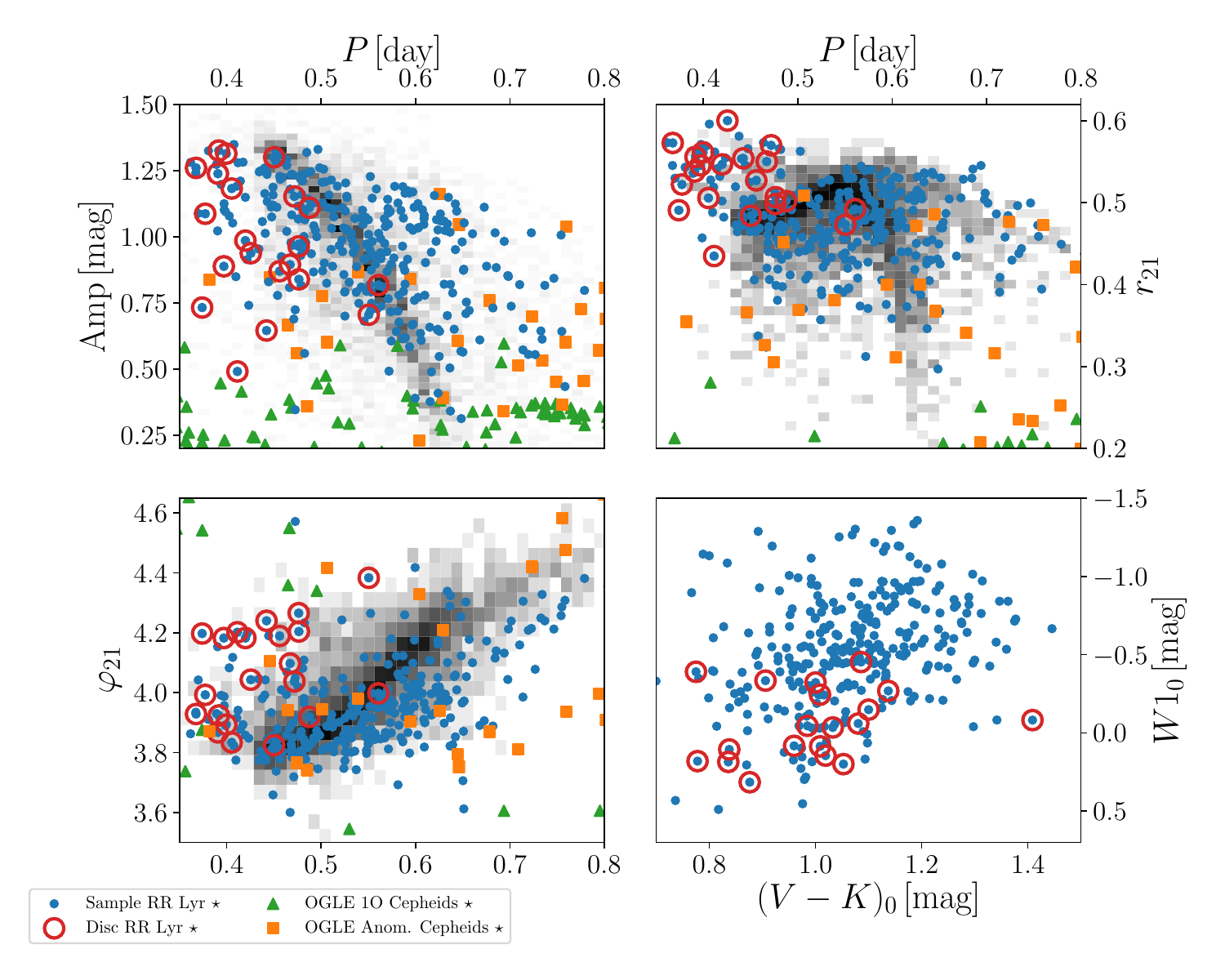}
\caption{A mosaic of pulsation properties and a colour-magnitude diagram (right-hand bottom panel) for our sample of RR~Lyrae stars represented by the blue points. Our pericenter peak RR~Lyrae variables are marked with red circles. The two top panels depict the period-amplitude diagram (left-hand panel) and amplitude ratio $r_{\rm 21}$ vs. pulsation period (right-hand panel). The bottom left-hand panel shows a phase difference $\varphi_{21}$ vs. pulsation period diagram. The bottom right-hand panel shows the colour-magnitude diagram for our sample of RR~Lyrae stars with the same colour-coding as in the previous three panels. In the three panels depicting the pulsation properties, the green triangles represent classical first overtone Cepheids from the Galactic bulge and Large Magellanic Cloud, and squares stand for anomalous Cepheids from the Large Magellanic Cloud \citep{Soszynski2015AnomCepheids}. The underlying gray distribution in the panels represents the pulsation parameters from the OGLE-IV $V$-band photometry \citep{Soszynski2014OGLEIV,Soszynski2017OGLEIV} for RR~Lyrae stars.}
\label{fig:PulsPropCMD}
\end{figure*}

We carefully examined the possibility that the pericenter peak RR~Lyrae stars are misclassified variables. Using the Fourier coefficients determined based on the ASAS-SN photometry, we compared the properties of the stars selected from Eq.~\ref{eq:FourierSeries} and \ref{eq:FourCoef} with other pulsating variable stars (Fig.~\ref{fig:PulsPropCMD}). 

For this comparison we used the $V$-band photometry from the fourth data release of the Optical Gravitational Lensing Experiment \citep[hereafter OGLE-IV,][]{Soszynski2014OGLEIV,Soszynski2017OGLEIV} for the Galactic bulge RR~Lyrae stars. The decision to use the bulge RR~Lyrae variables was based on their assumed metallicity \citep[similar to the metallicity of the pericenter peak RR~Lyrae stars $\text{[Fe/H]}=-1.02$\,dex,][]{Pietrukowicz2015}, and high-quality photometry for over 27\,000 fundamental mode RR Lyrae stars. The pericenter peak RR~Lyrae stars fall into the regions typical for the fundamental mode RR~Lyrae stars with short pulsation periods and asymmetric light curves. In the period-amplitude diagram they can be all clearly associated with the Oosterhoff type~I group \citep{Oosterhoff1939} identified in the Galactic bulge \citep{Prudil2019OO}. Moreover, a large fraction of the pericenter-peak RR~Lyrae stars occupy the regions of high-amplitude short-period \citep[from hereon referred to as HASP,][]{Fiorentino2015} RR~Lyrae stars, which are found mainly in the most metal-rich systems (cf. Fig.\ref{fig:FeV}). This further supports that their origin is intrinsic to the MW, since apart from the MW, HASP stars reside in metal-rich environments, and it is unlikely that a merger with a massive dwarf galaxy would result in circular orbits of the accreted stars. A colour-magnitude diagram (Fig.~\ref{fig:PulsPropCMD}, bottom-right panel) shows that the pericenter peak RR~Lyrae variables are slightly dimmer than the studied RR~Lyrae stars. This is mainly due to their high metallicity, which results in two effects in the stellar structure: an increase in the radiative opacity in the stellar atmosphere and a reduced mass of the helium core.

In the panels for the Fourier coefficients of Fig.~\ref{fig:PulsPropCMD}, we also included data for the first-overtone classical Cepheids, which can have a similar pulsation period as the fundamental mode RR~Lyrae stars. The classical Cepheids are Population I variables with ages up to a few hundred million years. We used the $V$-band photometry for the sample of classical Cepheids identified  by OGLE-IV in the Galactic bulge and Large Magellanic Cloud \citep{Soszynski2015Cepheids,Soszynski2017OGLEIV,Soszynski2017HenLeav}. Using the Eqs.~\ref{eq:FourierSeries} and \ref{eq:FourCoef}, we estimated their Fourier coefficients. In the period-amplitude diagram we see that classical Cepheids have similar pulsation periods as our pericenter peak RR~Lyrae stars, but much lower amplitudes. In addition, the $r_{\rm 21}$ vs. pulsation period dependence shows that the identified pericenter peak RR~Lyrae stars have very asymmetric light curves, unlike classical Cepheids. In general, we see that pericenter peak RR~Lyrae stars deviate from the general sample population in all four panels of Fig.~\ref{fig:PulsPropCMD}.

We performed a similar test using the anomalous Cepheids observed by OGLE-IV in the Large Magellanic Cloud \citep{Soszynski2015AnomCepheids}. They are metal-deficient helium-burning variables with higher masses than RR~Lyrae stars \citep[between $1.3-2.2$\,$\rm M_{\odot}$,][]{Marconi2004}, with uncertain origin\footnote{They are believed to be either intermediate-age variables \citep[1 - 6\,Gyr,][]{Norris1975,Demarque1975} or a product of mass transfer in binary systems with ages above 10\,Gyr \citep{Renzini1977}.}. They occur usually in nearby dwarf galaxies, but very rarely in globular clusters \citep[see][]{Kinemuchi2008, Bernard2009}. In the panels for pulsation properties, we see that anomalous Cepheids occupy similar regions as our pericenter peak RR~Lyrae stars. But the major difference between anomalous Cepheids and our pericenter peak RR~Lyrae stars is in the absolute magnitudes where anomalous Cepheids are intrinsically brighter ($M_{V} \approx (0.2; -1.5)$\,mag), while the majority of the pericenter peak RR~Lyraes are no brighter than 0.45\,mag. The exceptions are AR Per and SW Cru. Their absolute magnitude in the $V$-band is higher than $-0.7$\,mag, and therefore brighter than 95\,\% of our entire sample. On the other hand, both stars do not stand out in their absolute magnitudes in the $K_{s}$, and $W1$ passbands, where they fall on the faint end of the absolute magnitude distribution. In addition, both variables lie close to the Galactic plane, $\left | \text{b} \right | < 2.5$\,deg, thus we believe that the correction for the extinction is responsible for this discrepancy. Therefore, based on this simple comparison, we presume that our pericenter peak variables are truly RR~Lyrae stars.

We note that we did not compare our stars with other variables, e.g. eclipsing binaries, spotted stars, non-radial pulsators, Mira variables, and $\delta$~Scuti pulsators. Since our pericenter peak RR~Lyrae stars occupy the instability strip in the CMD (see the bottom right-hand panel of Fig.~\ref{fig:PulsPropCMD}) and exhibit asymmetric light curves without any sign of additional radial or non-radial modes we believe that they belong to none of the aforementioned variable classes. Some of the pericenter peak RR~Lyrae stars even exhibit a modulation of their light curves \citep[6 out of 22 stars are marked as modulated in the database of Blazhko stars\footnote{The online version can be found here \url{https://www.physics.muni.cz/~blasgalf/}},][]{Skarka2013}, which points toward the Blazhko effect \citep{Blazhko1907}, a typical phenomenon observed among all RR~Lyrae subtypes\footnote{For more information about the Blazhko effect see \cite{Szabo2014} and \cite{Catelan2015book}.}. Moreover, in the period-amplitude diagram, they lie in a region occupied by the fundamental-mode RR~Lyrae stars. 

Another explanation of their disc association can be erroneous proper motions due to their binary nature. Binary systems with an RR~Lyrae component are very rare; to this day only TU~UMa seems to be bound in a binary system \citep[e.g.,][]{Wade1999,Liska2016a}. Another 105 RR~Lyrae stars are potential candidates for a binary system, based on the RRLyrBinCan database\footnote{The online version of the database can be found here: \url{https://rrlyrbincan.physics.muni.cz/}} \citep{Hajdu2015,Liska2016Data,Liska2016b,Prudil2019Binary}. From our 22 RR~Lyrae stars, four (RS~Boo, XZ~Dra, CN~Lyr, DM~Cyg) are considered to be binary candidates and their proper motion could have been affected by the unknown second component. For an assessment of the astrometric solution for the studied stars, we can use the RUWE parameter provided in the {\it Gaia} DR2. Of the aforementioned 4 RR~Lyrae stars only DM~Cyg has a RUWE parameter higher than 1.4 -- (1.5, respectively). The RUWE parameter of the remaining three stars is below 1.2, therefore a reasonably good astrometric solution, hence these stars are most likely not affected by a possible second object in their system.

For these reasons, we consider our disc variables as fundamental-mode RR~Lyrae stars. 

\section{Conclusions} \label{sec:Conclus}

We report the detection of a peak in the pericenter distance distribution in the local RR~Lyrae star sample. The majority of these stars 22 have kinematics, abundances and pulsation properties that link them to the Galactic disc. It is important to note that our sample does not contain RR~Lyrae stars found independently in the Galactic disc based on a spatial distribution study \citep[using the V\'ia L\'actea survey][]{Minniti2010VVV,Dekany2018} nor in the direction of the Galactic bulge \citep{Soszynski2014OGLEIV,Soszynski2017OGLEIV}. This may have, to some extent, affected the selection function of our sample and the observed peak in the pericenter distance distribution. For RR~Lyrae stars located close to the Galactic plane ($\left | \text{b} \right | < 5$\,deg), we would expect to have similar $z_{\text{max}}$ and $e$ but not necessarily the same $r_{\text{per}}$.

These 22 pulsators pass the various tests of their association with the Galactic disc, e.g., based on their angular velocities and metallicities \citep{Layden1996} and velocities in the local standard of rest \cite{Bensby2003}. They also show the orbital properties associated with the disc RR~Lyrae stars described earlier by \cite{Maintz2005}. The pericenter peak RR~Lyrae pulsators have an average angular velocity $v_{\theta} = 241$\,km\,s$^{-1}$ and an average [Fe/H]=$-0.60$\,dex. They also stand out in their distribution of velocity components and are clearly distinct from the halo RR~Lyrae stars in our sample as well as from the RR~Lyrae stars that we find to be associated with the {\it Gaia}-Enceladus \citep[or {\it Gaia} Sausage,][]{Belokurov2018,Helmi2018Nature}. For some of the studied variables, we were able to find additional chemical information about their [Ca/Fe] ratio, which serves as a proxy for the $\alpha$ abundance. Our stars fall in the region occupied by $\alpha$-poor metal-rich stars \citep[e.g.,][]{Edvardsson1993,Reddy2003,Bensby2003,Roederer2014}. Their chemical similarity in [Ca/Fe], as well as their very similar orbital parameters, suggest that the pericenter peak RR~Lyrae stars have a common origin. The RR~Lyrae stars associated with the {\it Gaia} Sausage, on the other hand, are located at the metal-poor end ([Fe/H] between $-1.2$ to $-1.8$\,dex) with a large metallicity spread while overlapping with the $\alpha$-poor population discovered by \cite{Nissen2010}.

The pericenter peak RR~Lyrae stars have an asymmetric drift equal to $-5$\,km\,s$^{-1}$, therefore similar to the asymmetric drift estimated for the thin disc stars \citep[e.g.,][]{Bensby2003,Golubov2013,Sysoliatina2018}. Moreover, the RR~Lyrae variables associated with the Galactic disc exhibit a total and vertical velocity dispersion equal to $\sigma_{\rm s} = 37$\,km\,s$^{-1}$, and $\sigma_{\rm z} = 16$\,km\,s$^{-1}$, respectively. Low velocity dispersion is expected for the kinematically cold thin-disc stars \citet[][see their fig.~4]{Hayden2017} or \citet[][see their fig.~17]{Casagrande2011}, although we note that the pericenter peak variables should have a higher velocity dispersion based on their age if they had typical RR~Lyrae ages $>$\,10\,Gyr.

The 22 RR~Lyrae stars seem to be clearly associated with the Galactic disc in the solar neighbourhood, based on their orbital and chemical properties for stars with available [Ca/Fe] abundance, and their pulsation characteristics rule out a confusion with the other variable stars such as classical and anomalous Cepheids. Their possible association with the thin disc (in the case of those with known [Ca/Fe]) is in agreement with the spatial and kinematical studies focusing on the RR~Lyrae stars identified in the Galactic disc \citep[e.g.,][]{Dekany2018,Marsakov2018}. To corroborate our results for the remaining pericenter peak RR~Lyrae stars we would need information about their $\alpha$-element abundances (e.g., [Ca/Fe] or [Mg/Fe]).

Generally, if we were dealing with stars of ages younger than $8$\,Gyr \citep{Haywood2013}, and with the same kinematics and chemical properties as our pericenter peak stars, we would assign them to the Galactic thin disc \citep[or $\alpha$-poor disc,][]{Adibekyan2013,Hayden2015,Bland-Hawthorn2019}. On the other hand, the ages of RR~Lyrae stars are well constrained between $10 - 13$\,Gyr, based on the youngest and oldest globular clusters in which RR~Lyrae stars were detected \citep{Catelan2009,VandenBerg2013}. This seems to be in contradiction to the studies of the stellar ages of the thin disc as derived by, e.g. \cite{Haywood2013} and many others. Theses boundaries are somewhat diffuse; we note that several stars in their sample, assigned to the metal-poor thin disc, have ages around 10\,Gyr \citep[see fig.~8 in][]{Haywood2013}. The pericenter peak RR~Lyrae stars then possibly belong to the chemical and kinematical distribution of the thick disc overlapping the region occupied by the thin disc stars. However, their [Fe/H] and [$\alpha$/Fe] values are consistent with young thin disc stars, which remains a conundrum \citep[see also the discussion in][]{Marsakov2018,Marsakov2019s,Marsakov2019m}.

We suggest that the 22 pericenter peak RR~Lyrae stars may be the metal-rich alpha-poor extreme of the Galactic thick disc RR~Lyrae population since our sample covers only $\approx$ 10\,\% of the local RR~Lyrae stars. Hopefully, with future spectroscopic surveys we can expand our sample and shed more light on the Galactic disc RR~Lyrae population.

\section*{Acknowledgements}

This work has made use of data from the European Space Agency (ESA) mission {\it Gaia} (\url{https://www.cosmos.esa.int/gaia}), processed by the {\it Gaia} Data Processing and Analysis Consortium (DPAC, \url{https://www.cosmos.esa.int/web/gaia/dpac/consortium}). Funding for the DPAC has been provided by national institutions, in particular the institutions participating in the {\it Gaia} Multilateral Agreement. Z.P. acknowledges discussions with U. Bastian, A. Just, and K. Sysoliatina. Z.P. acknowledges the support of the Hector Fellow Academy. E.K.G and I.D. were supported by Sonderforschungsbereich SFB 881 ``The Milky Way System'' (particulary subprojects A03, A05, A11) of the German Research Foundation (DFG). We thank the anonymous referees for useful comments, which helped to improve the paper.



\bibliographystyle{mnras}
\bibliography{biby}

\appendix
\section{Table of the calculated kinematical properties of our studied stars} \label{sec:AppTable}

\setlength{\tabcolsep}{3.pt}
\renewcommand{\arraystretch}{1.5}
\begin{landscape}
\begin{table}
\footnotesize

\caption{The photometric, chemical, and orbital properties of our studied RR~Lyrae stars. The first lines of the table are shown here for illustration. The full table can be found in the supplementary material. The first column contains identification names for individual stars. The second column provides the pulsation period. Columns 3, 4, 5, 6, 7, and 8 list the determined photometric properties of the studied stars (mean magnitudes, amplitudes, and Fourier parameters). Columns 9 and 10 contain chemical information about metallicity and calcium abundance of a given star. Columns 11, 12, and 13 list Galactocentric spherical velocity components. The orbital parameters are listed in columns 14, 15, 16, and 17. The identified possible thin-disc stars are marked with an asterisk behind their name in Column 1. Similarly, the RR~Lyrae variables possibly kinematicaly associated with the {\it Gaia}-Enceladus/Sausage are marked with a plus sign.}
\label{tab:KinProp} 
\begin{tabular}{l|c|ccccccccccccccc}
\hline
Name & Period & V & Amp & $R_{\rm 21}$ & $\varphi_{21}$ & $R_{\rm 31}$ & $\varphi_{31}$ & [Fe/H] & [Ca/Fe] & $v_{\rm R}$ & $v_{\theta}$ & $v_{z}$ & $r_{\rm per}$ & $z_{\rm max}$ & $e$ & $r_{\rm apo}$ \\
 & [day] & [mag] & [mag] & & & & & [dex] & [dex] & [km\,s$^{-1}$] & [km\,s$^{-1}$] & [km\,s$^{-1}$] & [kpc] & [kpc] & & [kpc] \\ \hline
ZZ-And	&	0.55454	&	13.06	$\pm$	0.03	&	0.93	&	0.51	$\pm$	0.01	&	3.87	$\pm$	0.03	&	0.35	$\pm$	0.01	&	1.85	$\pm$	0.04	&	-1.58	&	--	&	243$^{23}_{22}$	&	-169$^{31}_{35}$	&	-183$^{22}_{25}$	&	4.8$^{0.9}_{0.7}$	&	21$^{10}_{6}$	&	0.78$^{0.04}_{0.05}$	&	38$^{14}_{9}$	\\
Z-Mic	&	0.58692	&	11.60	$\pm$	0.04	&	0.61	&	0.46	$\pm$	0.02	&	4.12	$\pm$	0.04	&	0.29	$\pm$	0.01	&	2.18	$\pm$	0.06	&	-1.28	&	0.27	&	87$^{3}_{3}$	&	194$^{2}_{2}$	&	-13$^{3}_{4}$	&	4.6$^{0.1}_{0.1}$	&	0.9$^{0.1}_{0.1}$	&	0.32$^{0.01}_{0.01}$	&	8.79$^{0.02}_{0.02}$	\\
Z-CVn	&	0.65394	&	11.90	$\pm$	0.03	&	0.75	&	0.48	$\pm$	0.02	&	4.05	$\pm$	0.05	&	0.26	$\pm$	0.02	&	2.16	$\pm$	0.09	&	-1.98	&	--	&	-132$^{6}_{9}$	&	-103$^{15}_{20}$	&	130$^{8}_{8}$	&	4.8$^{0.8}_{0.7}$	&	8.6$^{0.4}_{0.3}$	&	0.40$^{0.05}_{0.03}$	&	11.2$^{0.9}_{0.5}$	\\
YZ-Aqr	&	0.55194	&	12.68	$\pm$	0.03	&	0.95	&	0.46	$\pm$	0.01	&	3.84	$\pm$	0.03	&	0.35	$\pm$	0.01	&	1.73	$\pm$	0.05	&	-1.55	&	--	&	61$^{4}_{4}$	&	239$^{9}_{9}$	&	187$^{9}_{10}$	&	7.43$^{0.02}_{0.01}$	&	9.3$^{1.2}_{0.9}$	&	0.36$^{0.04}_{0.04}$	&	16$^{2}_{1}$	\\
YY-Tuc$^{+}$	&	0.63489	&	11.96	$\pm$	0.04	&	1.15	&	0.54	$\pm$	0.01	&	3.98	$\pm$	0.02	&	0.34	$\pm$	0.01	&	1.92	$\pm$	0.03	&	-1.82	&	--	&	-22$^{3}_{3}$	&	-3$^{10}_{11}$	&	47$^{6}_{6}$	&	0.63$^{0.05}_{0.03}$	&	3.01$^{0.04}_{0.05}$	&	0.85$^{0.01}_{0.01}$	&	7.54$^{0.03}_{0.03}$	\\
Y-Lyr	&	0.50270	&	13.26	$\pm$	0.04	&	1.25	&	0.56	$\pm$	0.01	&	4.03	$\pm$	0.01	&	0.33	$\pm$	0.01	&	1.96	$\pm$	0.02	&	-1.03	&	--	&	-8$^{2}_{2}$	&	167$^{14}_{15}$	&	22$^{6}_{6}$	&	4.4$^{0.7}_{0.6}$	&	1.5$^{0.1}_{0.1}$	&	0.29$^{0.07}_{0.07}$	&	7.99$^{0.03}_{0.03}$	\\
XZ-Oct	&	0.47270	&	13.35	$\pm$	0.03	&	0.35	&	0.16	$\pm$	0.09	&	4.57	$\pm$	0.57	&	0.15	$\pm$	0.09	&	1.37	$\pm$	0.64	&	-1.76	&	--	&	-198$^{12}_{13}$	&	37$^{18}_{19}$	&	112$^{13}_{13}$	&	0.9$^{0.2}_{0.1}$	&	4.1$^{0.4}_{0.3}$	&	0.85$^{0.02}_{0.03}$	&	10.9$^{0.3}_{0.3}$	\\
XZ-Mic$^{+}$	&	0.44915	&	12.98	$\pm$	0.04	&	1.16	&	0.47	$\pm$	0.02	&	3.85	$\pm$	0.04	&	0.29	$\pm$	0.01	&	1.67	$\pm$	0.06	&	-1.22	&	--	&	-82$^{9}_{10}$	&	-46$^{21}_{23}$	&	95$^{8}_{8}$	&	0.8$^{0.4}_{0.2}$	&	3.3$^{0.3}_{0.2}$	&	0.79$^{0.05}_{0.08}$	&	7.1$^{0.1}_{0.2}$	\\
XZ-Dra$^{*}$	&	0.47650	&	10.24	$\pm$	0.09	&	0.84	&	0.51	$\pm$	0.04	&	4.20	$\pm$	0.11	&	0.28	$\pm$	0.04	&	2.05	$\pm$	0.18	&	-0.87	&	0.10	&	29.3$^{0.6}_{0.6}$	&	227.3$^{0.6}_{0.6}$	&	-32.2$^{0.5}_{0.5}$	&	7.39$^{0.02}_{0.02}$	&	0.70$^{0.01}_{0.01}$	&	0.093$^{0.001}_{0.002}$	&	8.91$^{0.04}_{0.03}$	\\
XZ-Cyg	&	0.46659	&	9.85	$\pm$	0.09	&	0.81	&	0.32	$\pm$	0.03	&	3.88	$\pm$	0.13	&	0.18	$\pm$	0.03	&	1.97	$\pm$	0.21	&	-1.52	&	0.25	&	39.9$^{0.6}_{0.6}$	&	166$^{5}_{5}$	&	-288$^{3}_{3}$	&	8.134$^{0.003}_{0.004}$	&	17.5$^{0.7}_{0.6}$	&	0.45$^{0.01}_{0.01}$	&	21.5$^{0.7}_{0.7}$	\\
XZ-Aps$^{+}$	&	0.58725	&	12.39	$\pm$	0.04	&	1.07	&	0.52	$\pm$	0.02	&	3.95	$\pm$	0.05	&	0.32	$\pm$	0.02	&	2.03	$\pm$	0.08	&	-1.57	&	0.30	&	-77$^{3}_{3}$	&	-37$^{5}_{5}$	&	143$^{8}_{7}$	&	1.0$^{0.3}_{0.1}$	&	6.0$^{0.3}_{0.7}$	&	0.77$^{0.02}_{0.06}$	&	8.04$^{0.02}_{0.01}$	\\
XY-Eri	&	0.55425	&	12.99	$\pm$	0.03	&	0.98	&	0.40	$\pm$	0.03	&	4.00	$\pm$	0.07	&	0.26	$\pm$	0.03	&	1.66	$\pm$	0.11	&	-2.08	&	--	&	197$^{9}_{8}$	&	122$^{4}_{5}$	&	-28$^{10}_{11}$	&	3.1$^{0.1}_{0.1}$	&	3.3$^{0.3}_{0.3}$	&	0.71$^{0.02}_{0.02}$	&	18.1$^{1.0}_{0.8}$	\\
XY-And	&	0.39872	&	13.75	$\pm$	0.03	&	1.09	&	0.46	$\pm$	0.03	&	3.89	$\pm$	0.06	&	0.28	$\pm$	0.02	&	1.71	$\pm$	0.11	&	-0.92	&	--	&	55$^{25}_{26}$	&	53$^{22}_{25}$	&	-43$^{18}_{18}$	&	1.3$^{0.6}_{0.5}$	&	2.0$^{1.0}_{0.5}$	&	0.79$^{0.09}_{0.08}$	&	10.5$^{0.6}_{0.4}$	\\
XX-Pup$^{+}$	&	0.51720	&	11.23	$\pm$	0.07	&	1.19	&	0.46	$\pm$	0.01	&	3.84	$\pm$	0.03	&	0.36	$\pm$	0.01	&	1.65	$\pm$	0.04	&	-1.42	&	--	&	263$^{3}_{3}$	&	-30$^{3}_{3}$	&	-68$^{4}_{4}$	&	0.7$^{0.1}_{0.1}$	&	4.1$^{0.4}_{0.2}$	&	0.93$^{0.01}_{0.01}$	&	19.0$^{0.5}_{0.5}$	\\
XX-Lib$^{+}$	&	0.69851	&	12.51	$\pm$	0.03	&	0.85	&	0.51	$\pm$	0.02	&	4.21	$\pm$	0.05	&	0.32	$\pm$	0.02	&	2.30	$\pm$	0.07	&	-1.47	&	--	&	44$^{10}_{9}$	&	-20$^{12}_{13}$	&	100$^{4}_{4}$	&	0.4$^{0.2}_{0.1}$	&	3.3$^{0.1}_{0.1}$	&	0.89$^{0.04}_{0.06}$	&	6.8$^{0.3}_{0.1}$	\\
XX-Hya	&	0.50775	&	12.05	$\pm$	0.05	&	1.11	&	0.44	$\pm$	0.02	&	3.89	$\pm$	0.05	&	0.35	$\pm$	0.02	&	1.66	$\pm$	0.06	&	-1.33	&	--	&	-261$^{13}_{13}$	&	31$^{15}_{15}$	&	-48$^{6}_{6}$	&	0.6$^{0.3}_{0.2}$	&	2.0$^{0.5}_{0.3}$	&	0.94$^{0.02}_{0.03}$	&	18$^{2}_{1}$	\\
XX-And$^{+}$	&	0.72276	&	10.71	$\pm$	0.09	&	1.01	&	0.48	$\pm$	0.01	&	4.21	$\pm$	0.04	&	0.36	$\pm$	0.01	&	2.29	$\pm$	0.05	&	-2.01	&	--	&	254$^{10}_{10}$	&	-86$^{16}_{16}$	&	-150$^{7}_{7}$	&	2.3$^{0.4}_{0.4}$	&	13.1$^{0.7}_{0.8}$	&	0.82$^{0.01}_{0.01}$	&	23$^{3}_{3}$	\\
X-Ret	&	0.49199	&	11.74	$\pm$	0.05	&	0.87	&	0.41	$\pm$	0.03	&	3.85	$\pm$	0.09	&	0.21	$\pm$	0.03	&	1.65	$\pm$	0.16	&	-1.32	&	--	&	269$^{8}_{9}$	&	24$^{7}_{6}$	&	6$^{6}_{7}$	&	0.4$^{0.1}_{0.1}$	&	2.6$^{0.4}_{0.4}$	&	0.95$^{0.01}_{0.01}$	&	16.9$^{0.8}_{0.8}$	\\
X-LMi	&	0.68435	&	12.37	$\pm$	0.03	&	1.01	&	0.53	$\pm$	0.01	&	4.12	$\pm$	0.03	&	0.33	$\pm$	0.01	&	2.11	$\pm$	0.04	&	-1.68	&	--	&	-155$^{8}_{9}$	&	101$^{8}_{9}$	&	8$^{9}_{10}$	&	2.4$^{0.2}_{0.2}$	&	2.7$^{0.4}_{0.3}$	&	0.69$^{0.03}_{0.03}$	&	12.7$^{0.5}_{0.4}$	\\
X-Crt$^{+}$	&	0.73284	&	11.41	$\pm$	0.04	&	0.64	&	0.50	$\pm$	0.02	&	4.25	$\pm$	0.06	&	0.30	$\pm$	0.02	&	2.48	$\pm$	0.08	&	-1.75	&	--	&	-159$^{7}_{7}$	&	-14$^{13}_{15}$	&	-111$^{10}_{12}$	&	0.5$^{0.2}_{0.2}$	&	5.5$^{1.0}_{0.7}$	&	0.92$^{0.03}_{0.04}$	&	11.4$^{0.6}_{0.2}$	\\
X-Ari$^{+}$	&	0.65117	&	9.70	$\pm$	0.09	&	0.88	&	0.47	$\pm$	0.04	&	3.61	$\pm$	0.10	&	0.23	$\pm$	0.04	&	1.71	$\pm$	0.18	&	-2.40	&	0.35	&	-24.6$^{0.6}_{0.6}$	&	-37$^{4}_{4}$	&	-16.7$^{0.8}_{0.8}$	&	0.7$^{0.1}_{0.1}$	&	0.48$^{0.02}_{0.01}$	&	0.85$^{0.02}_{0.02}$	&	8.64$^{0.01}_{0.01}$	\\
WZ-Hya$^{+}$	&	0.53772	&	10.85	$\pm$	0.09	&	0.89	&	0.48	$\pm$	0.02	&	3.94	$\pm$	0.04	&	0.29	$\pm$	0.02	&	1.92	$\pm$	0.07	&	-1.39	&	--	&	31.5$^{0.8}_{0.7}$	&	-47$^{2}_{2}$	&	113$^{2}_{2}$	&	1.33$^{0.03}_{0.03}$	&	5.3$^{0.1}_{0.2}$	&	0.74$^{0.01}_{0.01}$	&	8.70$^{0.01}_{0.01}$	\\
WY-Vir	&	0.60935	&	13.44	$\pm$	0.03	&	1.08	&	0.44	$\pm$	0.01	&	3.82	$\pm$	0.03	&	0.36	$\pm$	0.01	&	1.58	$\pm$	0.03	&	-2.84	&	--	&	-106$^{3}_{3}$	&	38$^{11}_{14}$	&	97$^{7}_{8}$	&	1.7$^{0.3}_{0.3}$	&	7.1$^{0.3}_{0.4}$	&	0.64$^{0.04}_{0.04}$	&	8.0$^{0.1}_{0.1}$	\\
WY-Scl$^{+}$	&	0.46369	&	13.21	$\pm$	0.03	&	1.27	&	0.48	$\pm$	0.02	&	3.86	$\pm$	0.04	&	0.33	$\pm$	0.02	&	1.66	$\pm$	0.05	&	-1.51	&	--	&	6$^{2}_{1}$	&	-12$^{18}_{22}$	&	-98$^{6}_{6}$	&	0.8$^{0.5}_{0.3}$	&	7.2$^{0.6}_{0.5}$	&	0.84$^{0.07}_{0.09}$	&	9.1$^{0.1}_{0.1}$	\\
WY-Pav$^{+}$	&	0.58858	&	12.11	$\pm$	0.03	&	0.62	&	0.48	$\pm$	0.01	&	4.26	$\pm$	0.03	&	0.27	$\pm$	0.01	&	2.50	$\pm$	0.05	&	-0.98	&	--	&	159$^{10}_{9}$	&	18$^{9}_{10}$	&	86$^{3}_{3}$	&	0.4$^{0.1}_{0.1}$	&	3.7$^{0.2}_{0.1}$	&	0.91$^{0.02}_{0.03}$	&	9.0$^{0.3}_{0.3}$	\\
WY-Dra	&	0.58894	&	12.68	$\pm$	0.03	&	1.16	&	0.51	$\pm$	0.02	&	3.96	$\pm$	0.07	&	0.34	$\pm$	0.03	&	2.14	$\pm$	0.09	&	-1.66	&	--	&	78$^{11}_{11}$	&	186$^{15}_{17}$	&	56$^{9}_{9}$	&	5.9$^{0.5}_{0.7}$	&	2.3$^{0.4}_{0.3}$	&	0.34$^{0.02}_{0.02}$	&	11.4$^{1.0}_{0.8}$	 \\
\dots & \dots & \dots & \dots & \dots & \dots & \dots & \dots & \dots & \dots & \dots & \dots & \dots & \dots & \dots & \dots  & \dots\\
\hline
\end{tabular}
\end{table}

\end{landscape}  


\bsp	
\label{lastpage}
\end{document}